\global\def\draftcontrol{0}

   \def\versionno{ topological orientifold }
\catcode`\@=11

\expandafter\ifx\csname draftcontrol\endcsname\relax\global\def\draftcontrol{0} 
\fi 

{\count255=\time\divide\count255 by 60 
\xdef\hourmin{\number\count255} 
\multiply\count255 by-60\advance\count255 by\time 
\xdef\hourmin{\hourmin:\ifnum\count255<10 0\fi\the\count255}} 
\def\draftdate{\number\month/\number\day/\number\year\ \ \ \hourmin } 

\newcommand\makepapertitle{\par

  \begingroup 
    \renewcommand\thefootnote{\@fnsymbol\c@footnote}% 
    \def\@makefnmark{\rlap{\@textsuperscript{\normalfont\@thefnmark}}}% 
    \long\def\@makefntext##1{\parindent 1em\noindent 
            \hb@xt@1.8em{% 
                \hss\@textsuperscript{\normalfont\@thefnmark}}##1}% 
     \newpage 
     \global\@topnum\z@   % Prevents figures from going at top of page. 
     \@makepapertitle 
     \thispagestyle{empty}\@thanks 
  \endgroup 
  \setcounter{footnote}{0}% 
  \global\let\thanks\relax 
  \global\let\makepapertitle\relax 
  \global\let\@makepapertitle\relax 
  \global\let\@thanks\@empty 
  \global\let\@author\@empty 
  \global\let\@date\@empty 
  \global\let\@title\@empty 
  \global\let\title\relax 
  \global\let\author\relax 
  \global\let\date\relax 
  \global\let\and\relax 
  \def\version{\let\version\@version\@gobble} 
} 
\def\@makepapertitle{% 
  \newpage 
   \ifnum\draftcontrol=1 {} 
   \version\versionno 
   \vskip 5.5em% 
   \else 
   \hfill\hbox to 3cm {\parbox{4cm}{\@pubnum}\hss}% 
   \vskip 6.5em% 
   \fi 
   \begin{center}% 
   \let \footnote \thanks 
      {\hskip -0\textwidth \hbox to 1\textwidth% 
        {\centerline{\Large\bf{\noindent\@title}}}}% 
     \vskip 2em% 
     {\normalsize%\large 
       \lineskip .5em% 
       \begin{tabular}[t]{c}% 
         \@author 
       \end{tabular}\par}% 
     \vskip 1.5em% 
     {\@bstract}% 
     \end{center}% 
     \vfill
     \@date%
     \vskip 1.5em%
   \par 
} 

\gdef\@pubnum{} 
\def\pubnum#1{% 
  \gdef\@pubnum{#1}} 

\gdef\@bstract{} 
\def\Abstract#1{% 
  \gdef\@bstract{% 
   \parbox{\textwidth-0pc}{% 
   \centerline{\bf Abstract}\penalty1000 
   \noindent%\abstractfont \baselineskip=12pt 
   \renewcommand\baselinestretch{1.0} 
   {#1}}} 
} 

\gdef\@email{}
\def\email#1{%
   \gdef\@email{%
   Email: {\tt #1}}
}

\def\ps@paper{\let\@mkboth\@gobbletwo% 
     \ifnum\draftcontrol=1 
        \def\@oddfoot{\hbox to \textwidth{\tiny \versionno \hfil\tiny\draftdate}% 
        \hskip -\textwidth \hbox to \textwidth{\hfil\rm\thepage\hfil}}% 
     \else\def\@oddfoot{\hbox to \textwidth{\hfil\rm\thepage\hfil}} 
     \fi 
     \let\@evenfoot\@oddfoot 
} 
\def\body{\clearpage 
          \pagestyle{paper} 
        } 
\newenvironment{acknowledgments}{% 
\vskip 3.25ex 
\addcontentsline{toc}{section}{Acknowledgments}
\noindent {\bf Acknowledgments} 
} 

\def\@version#1{\ifnum\draftcontrol=1 
\typeout{}\typeout{#1}\typeout{} 
\vskip3mm\centerline{\hbox{\fbox{\normalsize{\tt DRAFT -- #1 -- } 
                   {\draftdate}}}}\vskip3mm 
\fi} 
\let\version\@version 
\long\def\eqlabel#1{\ifnum\draftcontrol=1 
                    \tag@false  % there are some problems with multline without this 
                    \tag*{(\theequation) \hbox to -0.2cm{\hspace{0cm}\small{#1}\hss}} 
                    \refstepcounter{equation}  
                    \edef\@currentlabel{\theequation} 
                    \ltx@label{#1}          % use old LaTeX \label instead of new definition 
                    \else 
                    \label{#1} 
                    \fi 
                    } 
\let\st@bibitem\@bibitem 
\let\st@lbibitem\@lbibitem 
\ifnum\draftcontrol=1 
  \def\@bibitem#1{% 
    \st@bibitem{#1}\a@@label{#1}\ignorespaces} 
  \def\@lbibitem[#1]#2{% 
    \st@lbibitem[#1]{#2}\a@@label{#2}\ignorespaces} 
  \def\a@@label#1{% 
    \gdef\a@lab{\smash{\normalfont\small#1}} 
    \ifvmode 
      \if@inlabel 
        \global\setbox\@labels\hbox{% 
          \llap{\a@lab\let\a@lab\relax 
                \kern\@totalleftmargin\kern\marginparsep}% 
          \box\@labels}% 
      \fi 
    \fi} 
\fi 
\documentclass[12pt,letterpaper]{article} 
\usepackage{amsmath,bm,amsfonts,amssymb,array,calc,amsthm,rotating}
\usepackage{epsfig,psfrag} 
\usepackage{graphicx}
\usepackage{color}
\usepackage[colorlinks=false]{hyperref}
\renewcommand\baselinestretch{1.25} 
\renewcommand\section{\@startsection {section}{1}{\z@}% 
                                   {-3.5ex \@plus -1ex \@minus -.2ex}% 
                                   {2.3ex \@plus.2ex}% 
                                   {\normalfont\large\bfseries}} 
\renewcommand\subsection{\@startsection{subsection}{2}{\z@}% 
                                   {-3.25ex\@plus -1ex \@minus -.2ex}% 
                                   {1.5ex \@plus .2ex}% 
                                   {\normalfont\normalsize\bfseries}} 
\renewcommand\subsubsection{\@startsection{subsubsection}{3}{\z@}% 
                                   {-3.25ex\@plus -1ex \@minus -.2ex}% 
                                   {1.5ex \@plus .2ex}% 
                                   {\normalfont\normalsize\it}} 
\renewcommand\paragraph{\@startsection{paragraph}{4}{\z@}% 
                                   {-3.25ex\@plus -1ex \@minus -.2ex}% 
                                   {1.5ex \@plus .2ex}% 
                                   {\normalfont\normalsize\bf}} 
\renewcommand\subparagraph{\@startsection{subparagraph}{5}{\z@}% 
                                   {-1.25ex\@plus -1ex \@minus -.2ex}% 
                                   {0ex \@plus .2ex}% 
                                   {\normalfont\normalsize\it}}

\numberwithin{equation}{section}

\long\def\@makecaption#1#2{%
  \vskip\abovecaptionskip
  \sbox\@tempboxa{{\bf #1:} #2}%
  \ifdim \wd\@tempboxa >\hsize
    {\small\bf #1:} {\small #2}\par
  \else
    \global \@minipagefalse
    \hb@xt@\hsize{\hfil\box\@tempboxa\hfil}%
  \fi
  \vskip\belowcaptionskip}
\renewcommand*\l@section[2]{%
  \ifnum \c@tocdepth >\z@
    \addpenalty\@secpenalty
    \addvspace{.5em \@plus\p@}%
    \setlength\@tempdima{1.5em}%
    \begingroup
      \parindent \z@ \rightskip \@pnumwidth
      \parfillskip -\@pnumwidth
      \leavevmode \bfseries
      \advance\leftskip\@tempdima
      \hskip -\leftskip
      #1\nobreak\hfil \nobreak\hb@xt@\@pnumwidth{\hss #2}\par
    \endgroup
  \fi}
\renewcommand*\l@subsection{\addvspace{.0em \@plus\p@}\@dottedtocline{2}{1.5em}{2.3em}}
\renewcommand*\l@subsubsection{\addvspace{-.2em \@plus\p@}\@dottedtocline{3}{3.8em}{3.2em}}

\def\hepth#1{\href{http://xxx.arxiv.org/abs/hep-th/#1}{{arXiv:hep-th/#1}}}
\def\mathph#1{\href{http://xxx.arxiv.org/abs/math-ph/#1}{{arXiv:math-ph/#1}}}

\def\math#1{\href{http://xxx.arxiv.org/abs/math/#1}{{arXiv:math/#1}}}

\def\mathsg#1{\href{http://xxx.arxiv.org/abs/math.SG/#1}{{arXiv:math.sg/#1}}}
\def\mathag#1{\href{http://xxx.arxiv.org/abs/math.AG/#1}{{arXiv:math.ag/#1}}}
\def\alggeom#1{\href{http://xxx.arxiv.org/abs/alg-geom/#1}{{arXiv:alg-geom/#1}}}
\def\arxiv#1#2{\href{http://xxx.arxiv.org/abs/#1}{{arXiv:#1 [#2]}}}

\definecolor{refcol}{rgb}{0.2,0.2,0.8}
\definecolor{eqcol}{rgb}{.6,0,0}
\definecolor{purple}{cmyk}{0,1,0,0}

\gdef\@citecolor{refcol}
\gdef\@linkcolor{eqcol}
\def\colorlinkspurple{\gdef\@urlcolor{purple}}
\def\colorlinksblue{\gdef\@urlcolor{blue}}
\def\colorlinksred{\gdef\@urlcolor{red}}

\def\ie{{\it i.e.}} 
\def\eg{{\it e.g.}} 
\def\etc{{\it etc.}}
\def\cf{{\it cf.}}

\def\revise#1       {\raisebox{-0em}{\rule{3pt}{1em}}% 
                     \marginpar{\raisebox{.5em}{\vrule width3pt\ 
                     \vrule width0pt height 0pt depth0.5em 
                     \hbox to 0cm{\hspace{0cm}{% 
                     \parbox[t]{4em}{\raggedright\footnotesize{#1}}}\hss}}}}

\newcommand\nxt[1]  {\\\fnxt#1} 

\def\cala         {{\cal A}} 
 
\def\calc         {{\cal C}} 
 
\def\cale         {{\cal E}} 
\def\calf         {{\cal F}} 
\def\calg         {{\cal G}} 
\def\calh         {{\cal H}}

\def\calk         {{\cal K}} 
\def\call         {{\cal L}} 
\def\calm         {{\cal M}} 
\def\caln         {{\cal N}} 
\def\calo         {{\cal O}}

\def\calr         {{\cal R}} 
 
\def\calt         {{\cal T}}

\def\calw         {{\cal W}}

\def\complex      {{\mathbb C}} 
 
\def\projective   {{\mathbb P}} 
 
\def\reals        {{\mathbb R}} 
\def\zet          {{\mathbb Z}} 
\def\CP{\complex\projective}
\def\RP{\reals\projective}

\def\del          {\partial} 
\def\delbar       {\bar\partial} 
\def\ee           {{\it e}} 
\def\ii           {{\it i}} 
 
\def\tr           {{\rm Tr}} 
 
\def\Im           {{\rm Im\hskip0.1em}}

\newcommand\topa[2]{\genfrac{}{}{0pt}{2}{\scriptstyle #1}{\scriptstyle #2}}

\def\sqr#1#2{{\vcenter{\vbox{\hrule height.#2pt   
 \hbox{\vrule width.#2pt height#1pt \kern#1pt 
 \vrule width.#2pt}\hrule height.#2pt}}}}

\def\ib{{\bar i}}
\def\jb{{\bar j}}
\def\kb{{\bar k}}
\def\lb{{\bar l}}
\def\mb{{\bar m}}

\def\0b{{\bar 0}}

\def\F#1#2{{\calf}^{(#1,#2)}}
\def\Fc#1{{\calf}^{(#1)}}
\def\K#1#2{{\calk}^{(#1,#2)}}
\def\R#1#2{{\calr}^{(#1,#2)}}
\def\G#1{{\calg}^{(#1)}}
\def\annulus{\cala}
\def\klein{\calk}
\def\moebius{\calm}

\def\Cu{{C^P}}
\def\Su{{S^P}}

\def\ch{{\rm ch}}

\def\Ipp{\mathord{\mathchar "0271 \kern-4.5pt \mathchar"0271}}

\def\TT{{\mathbb T}}
\def\aut{\mathop{\rm Aut}}

\def\eul{{\bf e}}

\catcode`\@=12 

\begin{document} 

%%% 
%%%%%% text starts here 
%%%%%%%%% 

\title{Evidence for Tadpole Cancellation in the Topological String}

\pubnum{%
arXiv:0712.2775}
\date{December 2007}

\author{
Johannes Walcher \\[0.2cm]
\it School of Natural Sciences, Institute for Advanced Study\\
\it Princeton, New Jersey, USA
}

\Abstract{
We study the topological string on compact Calabi-Yau threefolds 
in the presence of orientifolds and D-branes. In examples, we find
that the total topological string amplitude admits a BPS expansion
only if the topological charge of the D-brane configuration is equal 
to that of the orientifold plane. We interpret this as a manifestation of
a general tadpole cancellation condition in the topological string that 
is necessary for decoupling of A- and B-model in loop amplitudes. Our 
calculations in the A-model involve an adapted version of existing 
localization techniques, and give predictions for the real enumerative 
geometry of higher genus curves in Calabi-Yau manifolds. In the B-model, 
we introduce an extension of the holomorphic anomaly equation to unoriented 
strings.
}

\makepapertitle

\body

\version\versionno

\vskip 1em

\tableofcontents
\newpage

\section{Introduction}

In this paper we continue, and in a sense complete, the program of
extending the BCOV \cite{bcov1,bcov2} computation of perturbative topological 
string amplitudes on {\it compact} Calabi-Yau manifolds to the open 
string sector. The main novelty is the realization that the inclusion of 
non-orientable worldsheets appears unavoidable for a satisfactory physical 
interpretation of the open topological string (at least beyond 
tree-level), on compact Calabi-Yau manifolds. 

There is of course no consistency condition that would be more familiar 
to string theorists than the cancellation of anomalies in the ten-dimensional
superstring \cite{grsch}. As is well-known \cite{polcai}, the potential 
anomalies can be seen to originate from tadpoles of unphysical, BRST trivial 
states from the Ramond-Ramond sector. These anomalies and the associated infinities 
can only be removed by ensuring that the Ramond-Ramond tadpoles are canceled 
at tree-level. This is in distinction to NS-NS sector tadpoles, which can
in principle be adjusted by a Fischler-Susskind mechanism. The general
principle, outlined in \cite{fms}, is that anomalies in string theory can 
always be thought of as arising from boundary (surface) terms on the moduli 
space in the verification of decoupling of BRST trivial states from loop 
amplitude computations.

Similar comments apply to modern constructions using the type I/II superstring, 
whenever the space transverse to the D-branes and orientifold planes is 
compact \cite{polgimon}. In supersymmetric situations, orientifold planes
are the only known sinks of Ramond-Ramond charge, which the D-brane
configuration has to be adjusted to match. In this way, tadpole/anomaly 
cancellation remains at the center of the idea that a consistent coupling to 
a fundamental theory of quantum gravity restricts the possible gauge and 
matter content to a finite set of possibilities, uniquely realized in 
string theory.

Given its central importance in the superstring, it has always seemed 
natural to ask whether tadpole cancellation has an analogue in the 
topological phase of string theory \cite{tst}. The topological string
otherwise shares many features with its more physical counterparts and
is moreover related to them by many more or less direct links. That
tadpole cancellation indeed plays a role in the topological string
was anticipated in the previous paper \cite{openbcov}, in the
sense that loop amplitudes involving open strings are only well-defined
in sectors in which the total topological D-brane charge vanishes.%
\footnote{Warning: In the context of the Fukaya category, the
obstruction to defining Floer homology is also sometimes referred
to as a ``tadpole''. This phenomenon however can be dealt with by a 
shift of the open string background, and is similar to the NS-NS tadpoles 
mentioned above. The tadpoles of concern in the present paper are
more fundamental, and cannot be removed by a shift of background.} 
In this paper, we will find further corroboration of this statement. 
Moreover, we will analyze the possibility of canceling the tadpoles 
using orientifolds, which was also mentioned in \cite{openbcov}. The
precise statement and implications will be clear by the end of the 
introduction.

At this point, we do not have a microscopic understanding of the tadpole
cancellation condition, although we can give a rough sketch of its possible 
worldsheet origin. Recall that the gauge algebra of the topological string 
coincides with that of the bosonic string, and originates from the topological
twist of an underlying unitary $\caln=2$ superconformal field theory.
The identification between BRST operators, anti-ghosts, and ghost number
current on the one hand and the $\caln=2$ superconformal generators on the
other hand is as follows:
\begin{equation}
\eqlabel{identify}
\begin{array}{ccc}
(Q,\bar Q) &\leftrightarrow &(G^+,\bar G^+) \\
(b_0,\bar b_0) &\leftrightarrow &(G^-,\bar G^-) \\
(bc,\bar b\bar c) &\leftrightarrow &(J,\bar J)
\end{array}
\end{equation}
More precisely, the identification we have given in \eqref{identify} is
usually labeled as ``type B'', the associated topological string called
the B-model. The A-model arises from the identification in which $\bar G^+$
and $\bar G^-$ are exchanged on the RHS and $\bar J\to -\bar J$. A peculiarity 
of the topological string that distinguishes it from more conventional
bosonic string backgrounds, is the absence of a ghost field as adjoint
of the anti-ghost. In the underlying $\caln=2$ SCFT, worldsheet CPT 
conjugation relates $(G^+,\bar G^+)$ with $(G^-,\bar G^-)$. Using the
dictionary \eqref{identify}, this leads to the statement that the
anti-ghost (\ie, $(b_0,\bar b_0)$) cohomology is isomorphic to the BRST cohomology, 
and in particular non-trivial, in clear distinction to the 26-dimensional 
bosonic string. The model obtained from the identification $(Q,\bar Q)
\leftrightarrow (G^-,\bar G^-)$ is referred to as the anti-topological
B-model.

It was realized by Bershadsky, Cecotti, Ooguri and Vafa \cite{bcov1,bcov2}
that the non-trivial states of the anti-ghost cohomology do not decouple 
from the topological amplitudes. The failure originates from the boundary
of moduli space and can be viewed as an anomaly in accord with the above 
mentioned general principle. In distinction to gauge anomalies in the 
superstring, this so-called holomorphic anomaly is not fatal for the model. 
Instead, it can be expressed as a recursive derivative constraint on the
perturbative topological string amplitudes. In this way, it has become a 
central ingredient in the successful computation of topological string 
amplitudes in various situations over the years, as well as having various 
other interesting connections. 

The BCOV holomorphic anomaly equation was recently extended to the open 
string in \cite{openbcov}. Similar equations, which can in fact be understood 
as a special case of the equations of \cite{openbcov}, were obtained in 
\cite{marcos1,emo} from the study of matrix models. Earlier work on the 
holomorphic anomaly in the open string appears in \cite{bcov2,agnt2}, more 
recent work includes \cite{coy,bonelli,alim,konishi}. For the wavefunction 
interpretation of the open topological string along the lines of 
\cite{wittenwave}, see \cite{integrable,anv,newa}.

Another feature of the origin of the topological string in twisted $\caln=2$
SCFTs is that the {\it mixed BRST-anti-ghost} cohomology is non-trivial.
Indeed, changing the identification of \eqref{identify} to $(Q,\bar Q)
\leftrightarrow (G^+,\bar G^-)$ simply leads to the A-model on the same
Calabi-Yau manifold, which is generically non-trivial. It was shown in 
\cite{bcov2} that the A-model deformations decouple from the topological
B-model amplitudes. This argument was tailored to closed string topological
amplitudes, and must be reexamined in the presence of boundaries, for the
following reason.

It is a fundamental observation \cite{oooy} that the topological charges 
of D-branes from the B-model are carried by the A-model, and vice-versa. 
For instance, A-branes on a Calabi-Yau manifold are supported on 
Lagrangian submanifolds, and the associated middle-dimensional cycles are 
naturally observables in the topological B-model. By combining this observation 
with the statements from the previous paragraphs, it is natural to view the 
$(G^+,\bar G^-)$-cohomology as the topological string analogue of the compact 
RR gauge potentials in the superstring. And by further analogy, it is natural
to expect that the decoupling of A- and B-model, which is presumably violated 
in the presence of D-branes, will in fact be restored precisely when the
tadpoles are canceled, \ie, the total topological D-brane charge vanishes.

The evidence for tadpole cancellation that we will present in this paper 
involves the relation of the topological string to BPS state counting in 
M-theory \cite{gova,oova}. In more detail, we will proceed as follows. We will 
begin in section \ref{background} with a brief review of orientifold backgrounds 
of topological string. This will be useful later on when we cancel the tadpoles
using orientifolds, and will also establish some notation. We then turn to
some explicit A-model calculations in section \ref{Amodel}. The examples
we study include the quintic in $\projective^4$, the bicubic in $\projective^5$,
and the total space of $\calo_{\projective^2}(-3)$, where in each case
the Lagrangian brane is the real locus (with respect to the natural
complex conjugation on projective space). The method we will use is
essentially Kontsevich's localization calculus on the space of stable maps
\cite{kontsevich}, appropriately adapted to open and unoriented string
computations. This is similar to refs.\ \cite{katzliu,grza,mayr,diaconescu,bfm1,open,psw},
with certain new ingredients, and will lead to a computation of higher genus
open/unoriented Gromov-Witten invariants on the three spaces 
mentioned above (the method can certainly be extended to many other examples).
Strictly speaking, the requisite definitions in Gromov-Witten theory have not
been given yet, so our results depend on certain assumptions during the 
localization procedure, and our confidence that these assumptions are correct
depends on the overall consistency of the results, and on the agreement with
the B-model.

The surprise occurs when we plug these higher-genus open Gromov-Witten invariants
into the general multi-cover formula of \cite{gova,oova,lmv}. Namely, when these
formulas are applied naively, the resulting expansion coefficients are not 
integer, and hence {\it cannot} be interpreted in terms of the spectrum of 
BPS states of an M-theory compactification on the Calabi-Yau manifold. It
turns out, however, that when we {\it add the contributions from orientable and 
non-orientable worldsheets}, at fixed order in string perturbation theory, and 
then judiciously apply the multi-cover formulas of \cite{oova,lmv}, then
the resulting expansion coefficients are integer. The relative coefficient 
between orientable and non-orientable worldsheets is consistent with the 
identification of the D-brane charge of the orientifold plane in 
\cite{sinhav}. The BPS interpretation of topological string amplitudes on 
orientifolds was also studied in \cite{aahv,bfm2} for freely acting 
orientifolds on non-compact manifolds, where the effects that we discuss 
here play no role.

Thus girt with some explicit A-model results as benchmark, we turn in section 
\ref{formal} to formal developments to reproduce and complete these computations 
in the B-model. In particular, we write down an extension of the holomorphic
anomaly equation of \cite{bcov2,openbcov} to unoriented strings. We will
also see that the holomorphic anomaly equation for the total topological
string amplitudes, \cf, eq.\ \eqref{orderchi}, simplifies, and is in fact very
simply related to the extended holomorphic anomaly equation of \cite{openbcov}.

In section \ref{Bmodel}, we will then solve these generalized holomorphic
anomaly equations for the examples at hand. We will be able to fix the holomorphic 
ambiguities to match the A-model computations of section \ref{Amodel}, thus 
completing the mirror symmetric picture. The ability to fix the holomorphic
ambiguity is a non-trivial check if we have more A-model data than free
parameters in the B-model. In fact, there is a certain class of worldsheets 
for which we can extract the general form of the holomorphic ambiguity to all 
orders, in a way reminiscent of the results in \cite{eyon,emo,remodel}.

The results for the holomorphic ambiguities in section \ref{Bmodel} differ 
from those in \cite{openbcov}, which were obtained by a more naive application 
of the multi-cover formulas, and imposing integrality and low-degree vanishing 
of the putative BPS invariants.

Finally, we summarize and discuss other open issues in section \ref{conclusions}.

\section{Orientifold of Topological String}
\label{background}

An {\it orientifold background} in string theory is the result of gauging
\cite{dlp,horava1,sagnotti,horava2} a closed string background, say $M$, 
by the combined action of worldsheet parity $\Omega$ with an involutive target 
space symmetry $\sigma:M\to M$. We will denote the combined symmetry by $P$,
\begin{equation}
\eqlabel{parity}
P = \sigma\circ \Omega
\end{equation}
as well as various other incarnations of $P$ such as the corresponding 
operator acting on the space of string states. As mentioned in the 
introduction, simply modding out by $P$ will in general lead 
to inconsistencies due to massless tadpoles and anomalies. Tadpole
cancellation requires the inclusion of some D-brane configuration in the 
background, with a certain fixed total charge determined by the parity $P$.

One can equivalently think of an orientifold as a more conventional theory 
of strings propagating on the space $M/\sigma$, together with dynamics of 
unoriented strings localized around the geometrical fixed point loci of 
$\sigma$ (orientifold planes), as well as open string dynamics localized 
around the locations of the D-branes.

\subsection{Orientifolds of Calabi-Yau manifolds}

Compactifying the type IIA or type IIB string on a Calabi-Yau 3-fold
leads to an $\caln=2$ supersymmetric theory in four dimensions. When 
orientifolding this theory, the desire to preserve some ($\caln=1$)
supersymmetry imposes restrictions on the allowed involutions $\sigma$ 
by which to dress worldsheet parity. For example, if we consider the 
type IIA string on a Calabi-Yau $X$, and we let the action on non-compact 
spacetime be trivial, then $\sigma$ acting on $X$ should be an 
anti-holomorphic involution that reverses the K\"ahler form and maps 
the holomorphic three-form to its conjugate. The fixed point locus 
(O6-plane) wraps a special Lagrangian submanifold of $X$. In contrast to D-branes, 
orientifold planes do not carry any gauge degrees of freedom, so we do not
need to specify any bundle on top of the special Lagrangian. Such an 
involution is known as A-type orientifold. In the mirror dual Calabi-Yau 
manifold $Y$, the corresponding B-type involution should be a holomorphic 
involution of $Y$. The fixed point locus is a holomorphic submanifold, or 
more generally, a collection thereof, possibly of different dimensions.

From the worldsheet point of view, A- and B-type involution are of course
distinguished by the action on the $\caln=2$ supercharges, \ie, 
$G^\pm\to \bar G^\mp$ or $G^\pm \to \bar G^\pm$ for A- and B-type 
orientifold, respectively. As a consequence, when we consider twisting
of the $\caln=2$ worldsheet theory to the topological string, the A-type 
parity is consistent with A-twist, and B-type parity is consistent with 
B-twist. For the rest of this discussion, let us fix the B-model on 
Calabi-Yau $Y$ for definiteness.

The effects of the orientifold on the massless fields (moduli of the Calabi-Yau)
are discussed extensively in the literature. Without delving into details, 
we shall here make a few points that will be important later, and come 
back to others as we go along. A basic property of the parity is that it must
of course be compatible with the chiral ring structure, in a sense that
\begin{equation}
\eqlabel{asense}
P(\phi_a\phi_b) = P(\phi_a) P(\phi_b)
\end{equation}
where $\phi_a$, $\phi_b$ are elements of the $(c,c)$-ring. Moreover, chiral
fields of $U(1)$-charge $(q,\bar q)=(1,1)$ (the marginal fields) must have 
{\it odd} parity in order to survive the orientifold projection \cite{brho}. 
This is because the superspace measure for F-terms, $\int d\theta^+d\theta^-$, 
picks a minus sign under B-type parity $\theta^\pm\to\theta^\mp$.

As a consequence of these two properties, we can work out the action of
parity on the ground states from the vacuum bundle of special geometry.
Let us fix the action of $P$ on the unique RR ground state of $U(1)$-charge
$(-3/2,-3/2)$ (related by spectral flow to the identity in the $(c,c)$-ring)
to be $-1$ (a priori, we have an overall sign ambiguity in the action of
$P$). Then the parity of the ground states of $U(1)$-charge $(-1/2,-1/2)$ 
that will form the ``orientifold vacuum bundle'' is even, the parity of
the ground states of $U(1)$-charge $(1/2,1/2)$ is odd, \etc\ In particular, 
the topological metric is {\it odd} under worldsheet parity. For example,
for trivial action on the target space, the representation of $P$ on the
vacuum bundle takes the form
\begin{equation}
\eqlabel{ofform}
P = {\rm diag}(-1,\delta_i^j,-\delta_{\ib}^{\jb},1)
\end{equation}

In the formal developments, it will be convenient to reserve a notation for
the action of the chiral ring combined with parity on the Ramond ground
states from the vacuum bundle. 
\begin{equation}
{B_{ia}}^b |b\rangle:= P\circ \phi_i |a\rangle = {C_{ia}}^c P^b_c |b\rangle
\end{equation}
where $C_i=\bigl({C_{ia}}^b\bigr)$ is the representation of the chiral ring on 
the vacuum bundle, and $P=\bigl(P^b_c\bigr)$ the representation of the parity. 
We have
\begin{equation}
\eqlabel{convenient}
B_i = P C_i = - C_i P 
\end{equation}
where we used \eqref{asense} and the condition $P(\phi_i)=-\phi_i$ from 
the above discussion.

\subsection{Categorical digression}

What do orientifolds look like from the categorical point of view? (Some of 
the following comments are drawn from \cite{howa}, to which we refer for 
further details.) The category of B-branes is the derived category of coherent 
sheaves of Calabi-Yau $Y$.  A B-type orientifold is simply the data of an 
anti-automorphism of $D^b(Y)$, \ie, a functor $P:D^b(Y)\to D^b(Y)$, that 
reverses the direction of morphisms and whose square is isomorphic to the 
identity functor. The simplest such functor is just duality on $D^b(Y)$, 
but more complicated parities can be obtained by dressing with geometric 
involutions, twists by line bundles, or other non-trivial automorphisms. 

From this point of view, an orientifold looks very elementary, except maybe
if we ask the question for a classification of possible parity symmetries.
Things become more interesting if we ask for the categorical representation
of the notion of orientifold plane. Index theorems (in the space of open
strings between a brane and its parity image) provide a realization 
of the orientifold plane in the cohomology of $Y$, and thus allow at least 
the computation of the D-brane charge of the O-plane (modulo torsion).
However, for purposes of the topological string, we require more information.

For example, if we combine the formulas for orientifold superpotentials given
in \cite{aahv} with the general observations on D-brane superpotentials
on compact Calabi-Yau manifolds from \cite{mowa}, we learn that to get 
tree-level data for the orientifolded topological string, it is in general 
necessary to at least find a representation of the O-plane in algebraic 
K-theory (modulo torsion). We will not need such a notion for our examples 
here, but in our formal discussions, we will assume that the natural 
generalization of the results of \cite{aahv,mowa} hold. It would be 
interesting to analyze these questions further.

For another study of D-brane superpotentials in orientifolds from the 
categorical point of view, see \cite{diaconescu2}.

\subsection{Organization of perturbation theory}

The worldsheet of our string, $\Sigma$, is a real, two-dimensional manifold, 
which can have boundaries, is unoriented, and possibly non-orientable.
(Usually referred to as a Riemann surface, non-orientable ones are 
traditionally called Klein surfaces.) String perturbation theory 
is defined by integrating over the moduli space of conformal structures on 
$\Sigma$ the appropriate correlators of the two-dimensional worldsheet theory, 
and summing over all topological types of $\Sigma$.

\begin{figure}
\begin{center}
\psfrag{eq}{$\cong$}
\epsfig{height=3.6cm,file=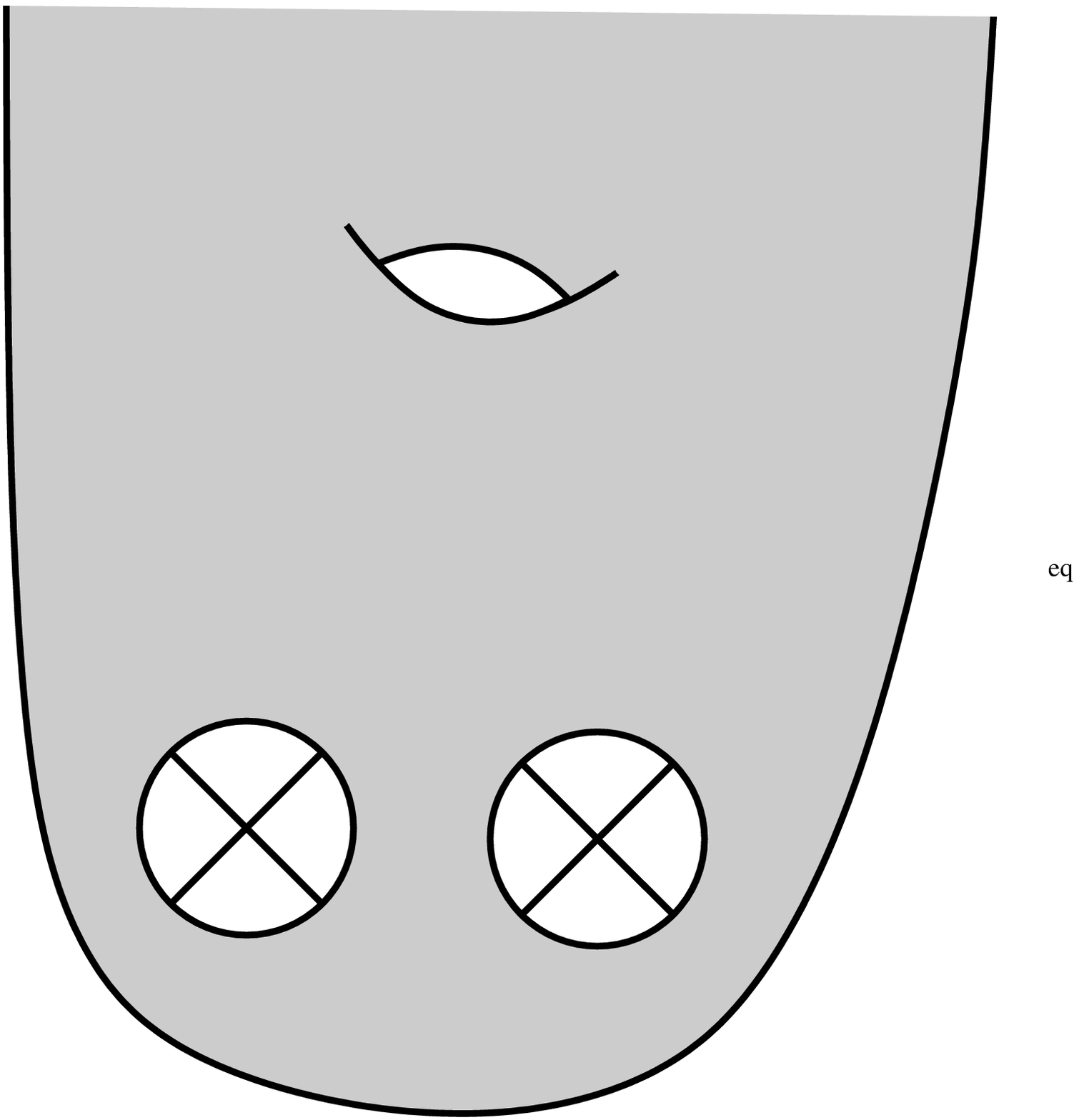}
\quad 
\epsfig{height=3.6cm,file=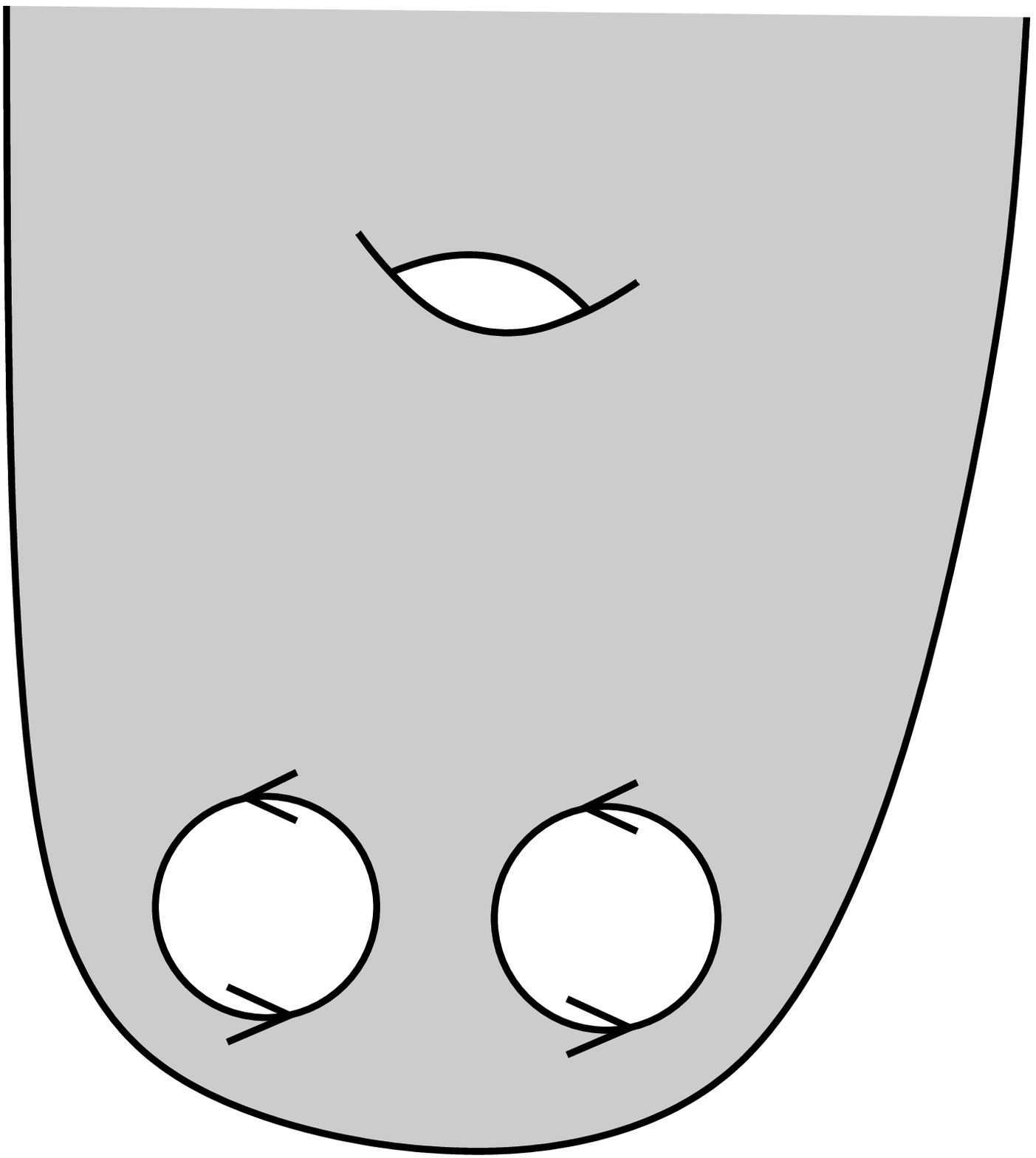}
\qquad
\epsfig{height=3.6cm,file=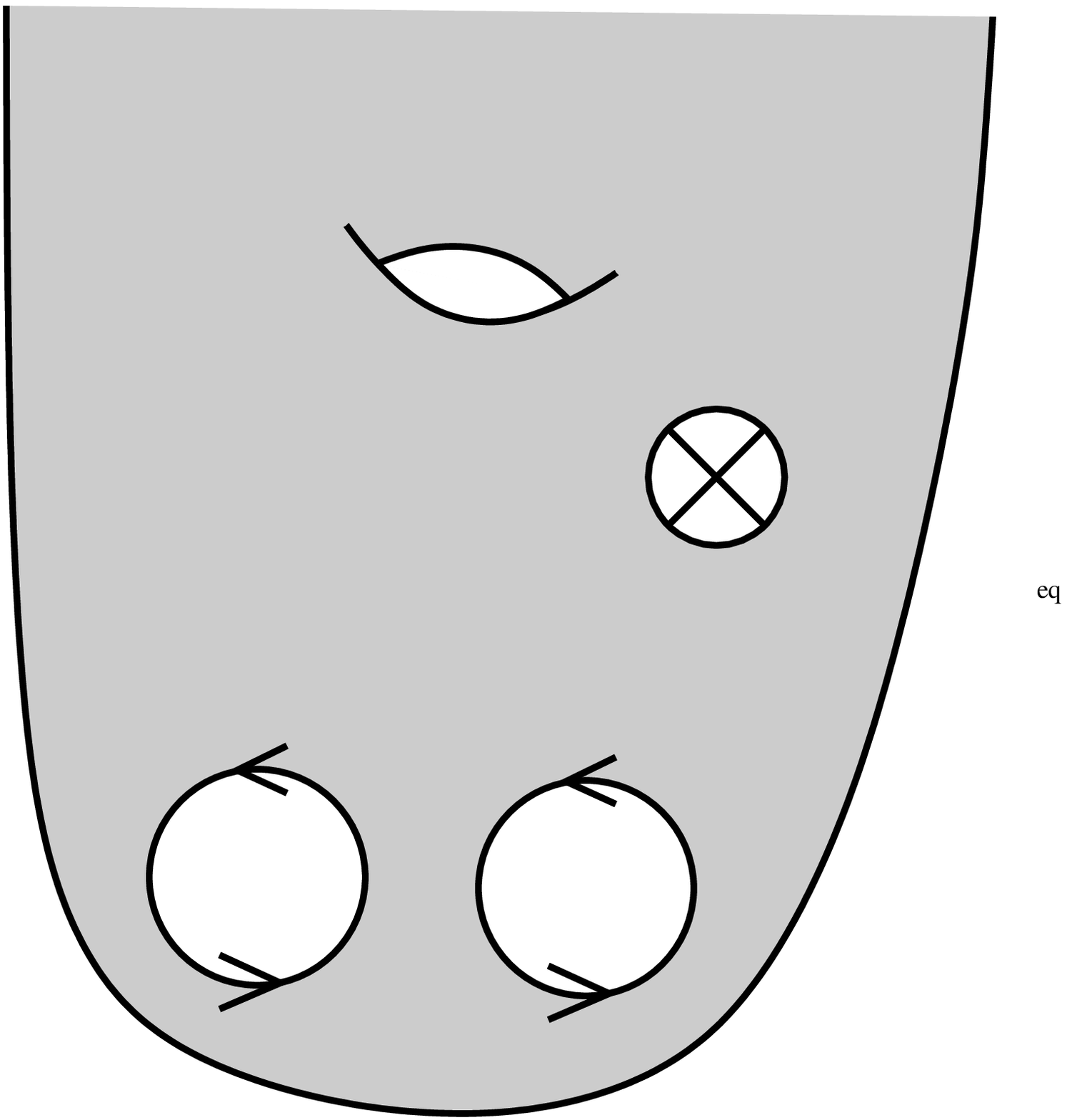} 
\quad
\epsfig{height=3.6cm,file=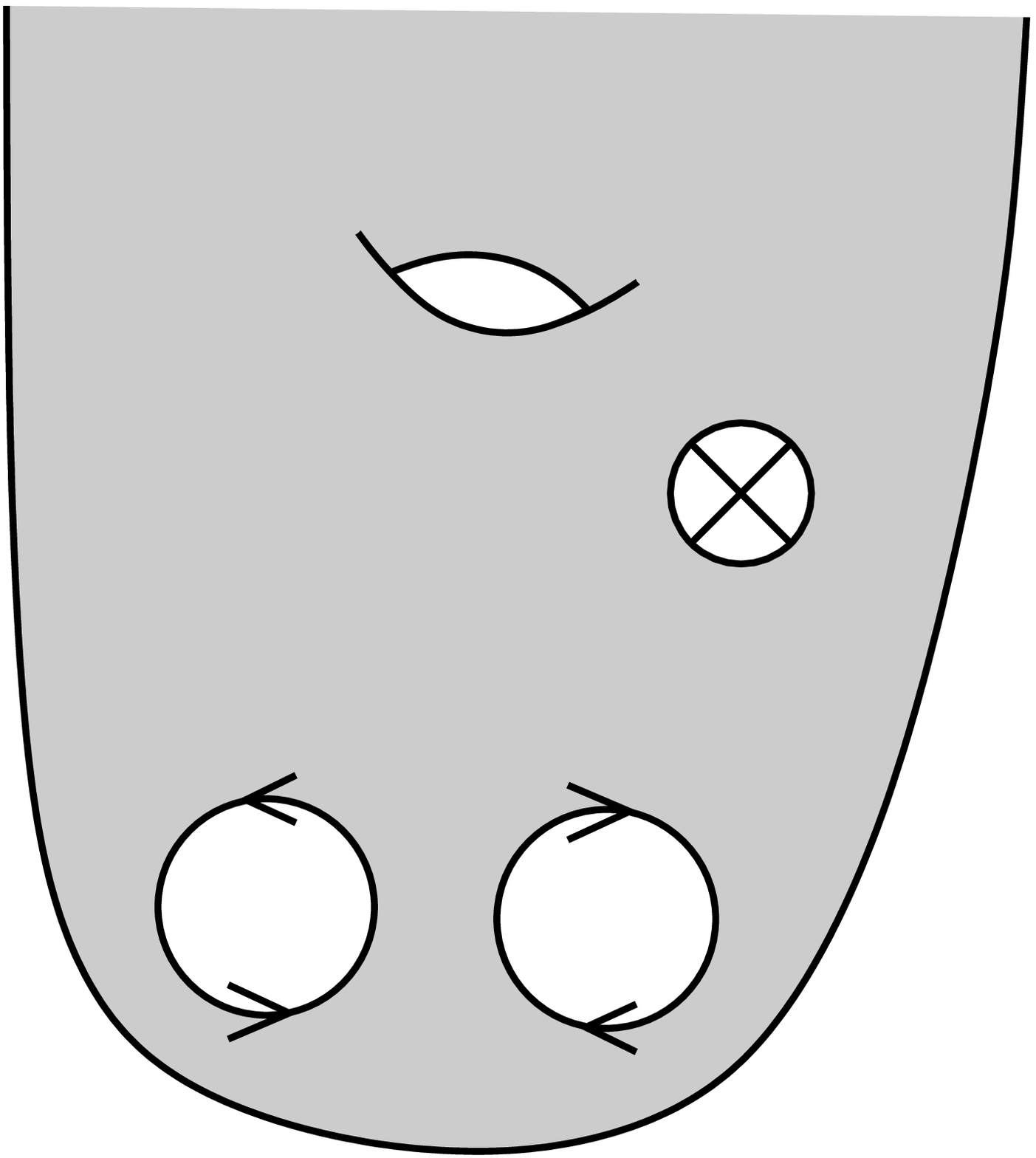} 
\caption{Locally on a Riemann surface, two crosscaps are equivalent to
a Klein handle, namely two holes glued together with an orientation 
reversal. In the presence of a third crosscap somewhere else on the Riemann 
surface, this is equivalent to an ordinary handle by pulling one of the
holes over the auxiliary crosscap.}
\label{equivalence}
\end{center}
\end{figure}

Smooth worldsheets are characterized topologically by the number of handles
(the genus) $g\ge 0$, the number of holes (boundary components) $h\ge 
0$, and the ``number'' of crosscaps, $c\ge 0$ that one attaches to the
standard Riemann sphere. Such a worldsheet contributes to string perturbation 
theory at the order determined by (the negative of) its Euler characteristic 
$\chi=2g+h+c-2$. Concerning $c$, one has to remember that three crosscaps can 
be traded (topologically) for one handle and one crosscap, as illustrated in 
Fig.\ \ref{equivalence}.

An alternative point of view on the various string worldsheets is
obtained by the so-called doubling construction. A Riemann surface $\Sigma$ 
as described above (with boundaries, possibly non-orientable) can be viewed 
as the quotient of an orientable Riemann surface without boundaries $\hat\Sigma$ 
by an orientation reversing involution,
\begin{equation}
\Sigma=\hat\Sigma/\Omega
\end{equation}
(Note that when $\Sigma$ itself is orientable and without boundaries, then
$\hat\Sigma$ has two connected components exchanged by $\Omega$. The mathematical
theory is more uniform, and more familiar, if we temporarily disregard this case.)
Inequivalent choices of $\Omega$ lead to different topologies for 
$\Sigma$. The conformal structure on $\Sigma$ induces a complex 
structure on $\hat\Sigma$ such that $\Omega$ becomes an anti-holomorphic 
involution. The pair $(\hat \Sigma,\Omega)$ where $\Omega$ is an 
anti-holomorphic involution of $\hat\Sigma$ is also known as a symmetric 
Riemann surface. We can think of the moduli space of conformal structures 
on $\Sigma$ as the moduli space of symmetric Riemann surfaces for this 
class of $\Omega$.

Let us consider a symmetric Riemann surface $(\hat\Sigma,\Omega)$, and denote
the genus of $\hat\Sigma$ by $\hat g$. The (negative) Euler 
character of $\Sigma$ is then given by
\begin{equation}
\chi = \hat g - 1
\end{equation}
The number of boundary components of $\Sigma$ is the number of
components of the fixed locus, $\Sigma_\Omega$, of $\Omega$. Finally, 
we can identify the index of orientability $k=0,1$ as $2-$(number of 
connected components of $\hat\Sigma \setminus \Sigma_\Omega$). These 
invariants are constrained by the conditions
\begin{equation}
\eqlabel{constrained}
\begin{array}{l}
\text{1. $0\le h \le \hat g+1$} \\
\text{2. For $k=0$ (orientable case), $h>0$ and $h\equiv \hat g+1\bmod 2$.}\\
\text{3. For $k=1$ (non-orientable case), $0\le h\le\hat g$.}\\
\end{array}\qquad\qquad
\end{equation}
Thus, for fixed $\hat g$, there are $\lfloor\frac{\hat g+2}2\rfloor$
orientable worldsheets that can be constructed as quotients $\hat\Sigma/\Omega$
and $\hat g+1$ non-orientable ones \cite{weichold} (a modern reference is, \eg, 
\cite{busep}). 

To make contact with string perturbation theory, we remember that $\hat g=\chi+1$,
and that for $\chi$ even, we also have the orientable worldsheet without boundaries
of genus $g_\chi=\frac{\chi}{2}+1$ (which is not of the form $\hat\Sigma/\Omega$
with $\hat\Sigma$ connected). Then the total number of worldsheet topologies 
that appear at order $\chi$ in perturbation theory is
\begin{equation}
\left\lfloor \frac{3\chi+8}2 \right\rfloor
\end{equation}
This count of course agrees with the one based on the description using
handles, boundaries and crosscaps. We have given the alternative
discussion here because it will be helpful further below to understand
the various ways in which the various Riemann surfaces can degenerate
when we vary the conformal structure. 

To organize the following discussion, we find it useful to separate 
the worldsheets into three classes, depending on whether $c$ is $0$, non-zero 
and odd, or non-zero and even. We will indicate this by the notation $\Sigma^{(g,h)}$,
$\Sigma^{(g,h)_r}$, and $\Sigma^{(g,h)_k}$, respectively. We shall denote 
the vacuum amplitudes on oriented surfaces by $\F gh$, those on non-orientable 
surfaces with an odd number of crosscaps by $\R gh$, and those on non-orientable 
surfaces with an even number of crosscaps by $\K gh$. \footnote{$\calf$ is for
Felix, and $\calk$ for Klein. $\calr$ could be for Riemann.} The precise
meaning of $g$ and $h$ in each case will become clear momentarily.

The total free energy of our topological string is then given by
\begin{equation}
\eqlabel{total}
\calg = \sum_\chi \lambda^\chi \G \chi
\end{equation}
where $\lambda$ is the string coupling, and $\chi$ the Euler character of 
the Riemann surface. $\calg^{(\chi)}$ is given according to the discussion 
above by
\begin{equation}
\eqlabel{orderchi}
\G \chi = \frac{1}{2^{\frac\chi 2+1}}\Bigl[
\calf^{(g_\chi)} 
+ \sum_{2g+h-2=\chi} \F gh 
+ \sum_{2g+h-1=\chi} \R gh
+ \sum_{2g+h-2=\chi} \K gh \Bigr]
\end{equation}
The first term in this sum is the purely closed string contribution
and appears for $\chi$ even, in which case $g_\chi=\frac{\chi}2+1$.

Before leaving this discussion, we add a remark from \cite{sinhav}
concerning the normalization of the topological string amplitudes. General
principles of string theory dictate that when we mod out by a symmetry group 
$G$ of order $|G|$, the amplitude of closed (and orientable) Riemann surface 
at genus $g$ contributes with a factor of $1/|G|^g$ compared to the original
theory before modding out. This is to ensure a consistent Hilbert space 
interpretation in the factorization of loop amplitudes. In our case,
$|G|=2$, and $g_\chi = \frac\chi 2+1$. This explains the prefactor in
\eqref{orderchi}, and we will confirm in section \ref{formal} that this
is indeed the correct normalization from the point of view of the holomorphic 
anomaly equation.

Note, however, that a prefactor of $2^{-\frac{\chi}2}$ can be reabsorbed in
\eqref{total} by redefining
\begin{equation}
\eqlabel{found}
\G \chi \to 2^{\frac \chi 2} \G \chi
\end{equation}
and the string coupling $\lambda\to\lambda/\sqrt{2}$. It was found in 
\cite{sinhav} that it is the normalization \eqref{found} of topological 
string amplitudes on the resolved conifold that allows for a natural comparison 
with results in the dual Chern-Simons theory on the deformed conifold.
We will confirm in this paper that this normalization is also the correct
one for the BPS expansion of topological string amplitudes on compact Calabi-Yau
manifolds. It would be interesting to understand this shift of the string
coupling more fundamentally.

\subsection{Remarks on moduli}

The topological amplitudes are functions (or rather, sections of an appropriate
bundle) over the moduli space of the underlying conformal field theory. When
A- and B-model are decoupled (in the presence of D-branes, we argue that
this requires canceling the tadpoles from the respectively ``other'' model), 
the amplitudes depend only on the K\"ahler moduli or complex structure moduli 
of the underlying Calabi-Yau manifold, respectively. (This dependence is not
holomorphic, due to the holomorphic anomaly.) For open strings, we also 
expect a dependence on open string moduli specifying positions and 
Wilson line degrees of freedom of the D-branes. A fundamental distinction
is that while infinitesimal closed string moduli are never obstructed,
open string moduli genuinely are, so it is a priori unclear how to
develop an analytic expansion as a function of these parameters. The open 
string moduli space can be expressed as the critical locus of the 
superpotential on some appropriate space of massive and massless 
(infinitesimal) deformations, varying over the moduli space of closed 
string deformations.

We emphasize again that open string moduli are not generically absent or physically 
irrelevant. (We don't want to argue against the important fact that the moduli 
space of the D0-brane at a point is just the Calabi-Yau manifold itself.) 
However, it was argued in \cite{openbcov} that, for {\it generic} 
values of the closed string moduli, topological 
string amplitudes {\it do not depend} on any continuous open string 
moduli, when such are present. We will naturally assume that this statement 
is true, and that the amplitudes only depend on the discrete moduli. Therefore, 
if we consider a D-brane configuration with $N$ branes, we will need $N$ 
discrete labels $v_1,\ldots v_N$ to specify it. (The total number of branes 
is not naturally fixed, although it is restricted if we impose tadpole 
cancellation with respect to some given orientifold plane.) The topological 
amplitudes can then be expanded as
\begin{equation}
\eqlabel{canbe}
\F gh (t;v_1,\ldots,v_N) = \sum_{i_1,\ldots i_h} \F gh_{v_{i_1},\ldots,v_{i_h}}(t)
\end{equation}
where $t$ denotes the closed string moduli. A similar expansion applies to the
non-orientable amplitudes, $\K gh$ and $\R gh$.

\section{A-model Computations}
\label{Amodel}

Localization on the moduli space of stable maps was originally introduced in
\cite{kontsevich} for the computation of genus $0$ Gromov-Witten invariants on the 
quintic, and the verification of physicists' mirror symmetry prediction \cite{cdgp}. 
Over the years, the method has also been successfully applied for the computation of 
all-genus (closed) Gromov-Witten invariants in local toric Calabi-Yau manifolds. 
More recently, the BCOV prediction of genus $1$ Gromov-Witten invariants of 
hypersurfaces was verified rigorously in \cite{zinger}, with localization 
as an important tool. In the case of local toric Calabi-Yau manifolds, 
with toric branes, localization was first used for the computation of 
open Gromov-Witten invariants in \cite{katzliu}, checking the prediction of 
\cite{oova}. The computations were extended to higher genus and multiple 
boundary components in \cite{grza,mayr}, and to freely acting orientifolds in 
\cite{diaconescu}. In \cite{open}, the method was adapted to the computation of 
open Gromov-Witten invariants on the quintic, with boundary conditions given 
by the real locus. In this situation, open Gromov-Witten invariants were defined 
in \cite{jake}, and the prediction of \cite{open} was then verified rigorously 
in \cite{psw}.

We will here combine these various parts, adding a few new ingredients. This 
will lead to a computation of open Gromov-Witten invariants on the quintic in 
genus $0$ with an arbitrary number of boundary components, and unoriented 
Gromov-Witten invariants in genus $1$, again with arbitrary number of 
boundary components. For local $\projective^2$, with a non-toric brane, 
we are able to compute these invariants for arbitrary worldsheet topology.

We remark that none of these new invariants (except genus $0$, with one boundary 
component) has yet been defined rigorously. As a consequence, we have to give some 
ad-hoc prescriptions how to deal with certain non-isolated fixed loci that 
occur in the localization process, as well as how to fix the signs related to 
the orientation of moduli spaces. The latter can probably be determined along the 
lines of \cite{jake,melissa}, while the former problem should be dealt with by 
methods similar to those used in \cite{psw}. We leave this for future research.

Let us emphasize one aspect of our results which marks a significant departure from the 
previous works we have mentioned. In those cases, the localization results actually 
depend on certain of the torus weights. This dependence is due to the need of choosing
certain boundary conditions in the space of stable maps \cite{melissa}, and can be
related to the framing ambiguity of knots in Chern-Simons theory \cite{agva,akv}. 
As a consequence, the resulting numbers are not truly invariants of the space with 
D-brane on top, although in all cases they do satisfy the integrality predictions of
\cite{oova,lmv}. In the cases of our interest, we find actual invariant, weight-independent 
localization results, which however do not satisfy integrality worldsheet by worldsheet.
Integrality is recovered once we sum over all worldsheet topologies at fixed order
in string perturbation theory. Summing over worldsheet topologies is related to
eliminating the boundaries in moduli space, which explains why we obtain an invariant
result. We will return to this in section \ref{formal}.

\subsection{The examples}

We consider three examples of Calabi-Yau manifolds, $X$: the quintic, the bicubic, and
local $\projective^2$. In the first two cases, we fix a particular choice of complex 
structure. Each of the three examples will be equipped with an anti-holomorphic involution
$\sigma:X\to X$. The fixed locus $L$ of $\sigma$ can either be turned into an A-brane
by specifying a flat line bundle on top of it, or it can arise as the O-plane in an
orientifold model based on $(X,\sigma)$. Later on, we will be interested in giving 
$L$ both roles simultaneously.

The A-model of course should not depend on the choice of complex structure. In particular,
once we equip $L$ with a flat line bundle, we obtain an object in the Fukaya category
of $X$ which remains invariant as we vary the complex structure of $X$. For example, 
the open Gromov-Witten invariants on the disk defined in \cite{jake} do not depend
on the complex structure, as discussed in \cite{psw}. However, it should be noted 
that once we deform away from the initial complex structure, $L$ need not be the 
fixed point of an anti-holomorphic involution any longer.

Once we orientifold, the complex structure is actually restricted to be invariant under
complex conjugation. Moreover, the topology and homology class of the orientifold plane
change along singular conifold loci in the moduli space. (See, \cite{bhhw,raul} for
a study of this phenomenon.) By combining this with the remarks from the previous 
paragraph, we learn that to maintain tadpole cancellation, new branes will have to 
be created when moving through these singular loci, as discussed in the superstring
context in \cite{raul}. It will be interesting to study the implications of this
for Gromov-Witten theory and BPS invariants.

More specifically, the examples are as follows. The quintic is the vanishing locus 
of a degree $5$ polynomial in $\projective^4$. Consider the Fermat case
\begin{equation}
\{ x_1^5+x_2^5+x_3^5+x_4^5+x_5^5=0\} \subset\projective^4
\end{equation}
This is invariant under complex conjugation $x_i\to\bar x_i$ on $\projective^4$. The 
fixed point locus $L=\{x_i=\bar x_i\}$ is topologically equal to $\reals\projective^3$, 
and admits two choices of flat line bundles. The corresponding discrete Wilson line 
will be denoted by $\epsilon=\pm 1$. For the localization computation, it is 
useful to change coordinates such that complex conjugation acts in a non-standard 
way
\begin{equation}
\eqlabel{nonstandard1}
\sigma:(x_1,x_2,x_3,x_4,x_5) \to (\bar x_2,\bar x_1,\bar x_4,\bar x_3,\bar x_5)\,,
\end{equation}
which is what we have in mind in the following.

The bicubic is the intersection of two cubic polynomials in $\projective^5$. For 
appropriate choice of complex structure, \eg,
\begin{equation}
\{x_1^3+x_2^3+x_3^3=0\} \cap \{x_4^3+x_5^3+ x_6^3 = 0 \}\subset\projective^5
\end{equation}
the real locus is also an $\RP^3$. Again, we switch to coordinates such 
that complex conjugation acts as
\begin{equation}
\eqlabel{nonstandard2}
\sigma:(x_1,x_2,x_3,x_4,x_5,x_6) \to (\bar x_2,\bar x_1,\bar x_4,\bar x_3,\bar x_6,\bar x_5)
\end{equation}

Finally, local $\projective^2$ is the total space of the canonical bundle over
the projective plane,
\begin{equation}
\calo_{\projective^2}(-3) \to \projective^2
\end{equation}
We take complex conjugation to act as
\begin{equation}
\eqlabel{nonstandard3}
\sigma:(x_1,x_2,x_3)\to (\bar x_2,\bar x_1,\bar x_3)
\end{equation}
on $\projective^2$ and by complex conjugation in the fiber. The real locus in this 
case is the total space of the orientation bundle over $\RP^2$. As before, 
$H_1(L;\zet)\cong\zet_2$, so we again have a choice of discrete Wilson line. 

For comparison with previous work, we will also evaluate our localization formulas
for the conifold, the total space of $\calo(-1)\oplus \calo(-1)$ over $\projective^1$.
The brane is the real locus which is isomorphic to the direct sum of two M\"obius 
strips over $\RP^1$, in other words $S^1\times\reals^2$. This is the brane originally 
studied in \cite{oova}. In distinction to the previous cases, $H_1(L;\zet)\cong\zet$. 
This entails further dependence of the localization results on certain torus weights,
as in \cite{katzliu}. To eliminate boundaries of moduli space and obtain an 
invariant result, we have to sum not only over worldsheet topologies, but also
over the boundary degree modulo $2$, as was done in \cite{psw}. This 
weight-independence of conifold results is similar to \cite{diaconescu}, although 
in that work the orientifold involution was taken to be freely acting.

\subsection{Localization}

Each of the three examples presented in the previous subsection comes with the
natural action of a torus $\TT^n$ on $\projective^{n-1}$, where $n=5,6,3$ for 
quintic, bicubic, and local $\projective^2$, respectively. The complex conjugation 
is compatible with a subtorus $\TT^{n'}$ with $n'=2,3,1$, respectively, which is 
most obvious when complex conjugation is taken to act as in \eqref{nonstandard1}, 
\eqref{nonstandard2}, \eqref{nonstandard3}. In each case, we have a canonical 
lift of the $\TT^n$ action to the bundles $\calo_{\projective^{n-1}}(k)$
compatible with the real structure. The conifold is invariant under $\TT^3$, and 
its real brane is compatible with a $\TT^2\subset\TT^3$. In this way, the real 
brane of the conifold is actually identical to one of the ``toric'' branes that 
have been much studied in the literature, starting with \cite{agva}.

Following the cited literature, we consider moduli spaces of maps of degree $d$ 
from a worldsheet $\Sigma$ of one of the topological types $(g,h)$, $(g,h)_r$, 
or $(g,h)_k$ discussed in section \ref{background} to the ambient target space
$\projective^{n-1}$, with some extra data, as follows.\footnote{We recall that 
the subscript $r$ indicates an odd number of crosscaps, while the subscript $k$ 
indicates that we have an even number of crosscaps. $h$ is the number of boundary 
components, and $g$ is such that the negative Euler characteristic is $2g+h-2$, 
$2g+h-1$ and $2g+h-2$ in the three cases respectively.} When $\Sigma$ is 
orientable with non-empty boundary, we require that the boundary be mapped 
to the real locus. Note that any such map can be completed to an invariant 
map from the doubled surface $\hat\Sigma$ to $\projective^{n-1}$. So more 
uniformly, whether $\Sigma$ is orientable or not, we can think of a map 
from the doubled worldsheet $\hat\Sigma\to \projective^{n-1}$ that is 
equivariant with respect to the action of $\Omega$ (where $\Sigma=\hat\Sigma/
\Omega$) on $\hat\Sigma$ and complex conjugation $\sigma$ on the target space. 
The fixed points of $\Omega$ are automatically mapped to the Lagrangian brane.
Since we have in mind a fixed boundary condition and orientifold projection in all 
cases, we denote the moduli space simply by
\begin{equation}
\eqlabel{three}
\calm_\Sigma(\projective^{n-1},d)
\end{equation}
and we reserve technicalities associated with the proper compactification of these
spaces for a later discussion.

We now come to an important point. As we just mentioned, any boundary of 
$\Sigma$ is automatically mapped to the Lagrangian $L$, but we have so
far not specified its homology class in $H_1(L)$, which is $\cong\zet_2$
in all three cases of interest.\footnote{The degree of the map specifies
the relative cohomology class in $H_2(X,L)$. This determines the total
class of the boundary in $H_1(L)$, but not of the individual boundary 
components.} When that class is trivial, then under deformation of 
the map (or under change of complex structure of the target space, 
for the quintic and bicubic), it can happen that the boundary is 
collapsed to a point on $L$. This is one of the real codimension-one 
boundaries in the moduli space that we have been warned about. As
we will see, it is the only dangerous one, at least in the examples we
consider here. From the doubled perspective, $\hat\Sigma$ develops a node 
that lies right on top of the Lagrangian $L$. Reflection shows that 
such a real nodal curve admits another smoothing to a curve that is 
equivariant, but with respect to a different anti-holomorphic involution 
of $\hat\Sigma$, that locally looks like a crosscap.

The local model of this phenomenon is the map
$\projective^1 \ni (u,v) \mapsto (x,y,z) \in \projective^2$ defined by
\begin{equation}
\eqlabel{onepar}
\begin{split}
x&= a u^2 \\
y&= a v^2 \\
z&= u v
\end{split}
\end{equation}
where $a$ is a parameter. The image of the map is the conic
\begin{equation}
\eqlabel{image}
xy-a^2 z^2 = 0
\end{equation}
Whenever $a^2\in\reals$, the image curve is real under $(x,y,z)\mapsto 
(\bar y,\bar x,\bar z)$, but for the map \eqref{onepar} to be equivariant, 
we have to choose the involution to act on $\projective^1$ as
\begin{equation}
\begin{split}
(u,v)& \mapsto (\bar v,\bar u) \;\;\; \qquad  a\in\reals \\
(u,v)& \mapsto (\bar v,-\bar u) \qquad  a\in \ii \reals
\end{split}
\end{equation}
In the first case, we obtain a map from the disk to $\projective^2$ with boundary
on $\RP^2$, in the second case we obtain a  map from the crosscap to the orientifold.
Thus, we see that as we vary the (target space) parameter $a^2$, we can have
transitions where we loose holomorphic disks and gain holomorphic crosscaps.
Since this is a local phenomenon, happening in real codimension $1$, the only
way to account for this process is to count disks with collapsible boundaries
and crosscaps together. Deferring a more careful discussion to a later stage,
we are now ready to explain how we will do the computations. We temporarily
assume that any boundary component is mapped to a non-trivial homology 
class in $H_1(L)$.

We intend to compute Gromov-Witten invariants, $\tilde n^\Sigma_d$, by 
integrating over the moduli space $\calm_\Sigma(\projective^{n-1},d)\ni f$ the 
top Chern class of an appropriate bundle $\cale_d$. In the three cases, we 
have
\begin{equation}
\cale_d = 
\begin{cases} H^0(\Sigma,f^*\calo(5)) & X={\rm quintic} \\
H^0(\Sigma,f^*\calo(3)^{\oplus 2}) & X={\rm bicubic} \\
H^1(\Sigma,f^*\calo(-3)) & X=\text{local $\projective^2$}
\end{cases}
\end{equation}
Namely, we assume that for each topological type of surface, we will have an
Euler class formula of the form
\begin{equation}
\eqlabel{euler}
\tilde n_d^\Sigma = \int_{\calm_\Sigma(\projective^{n-1},d)} \eul(\cale_d)
\end{equation}
which we will evaluate by using Atiyah-Bott localization following the cited 
literature.

As explained in \cite{kontsevich}, the fixed loci of the action of $\TT^n$ are 
nodal curves in which any node or any component of non-zero genus is collapsed 
to one of the fixed points in target space, and any non-contracted rational 
component is mapped on one of the coordinate lines with a standard map
of a certain degree. The components of the fixed locus can be represented 
by a decorated graph,
\begin{equation}
\Gamma= \{v_1,\ldots,v_k;e_1,\ldots, e_l;p_1,\ldots,p_k;g_1,\ldots,g_k;
d_1,\ldots, d_l\}
\end{equation}
The graph data consists of a set of vertices $(v_i)_{i=1,\ldots, k}$, and a set 
of edges $(e_j)_{j=1,\ldots, l}$. The decoration consists of the genus $g_i\ge 0$
of any contracted component at the $i$-th vertex, the target space fixed 
points $1\le p_i\le n$ to which that component is mapped, and the degree
$d_j$ of the maps from the edges to the corresponding coordinate line. We will 
conveniently omit the decorations $g_i=0$ and $d_j=1$.

We are here interested in localization with respect to $\TT^{n'}$ on the space 
of real maps, \ie, maps equivariant with respect to conjugation on target and 
worldsheet
\begin{equation}
\eqlabel{symmetric}
(\hat\Sigma,\Omega) \to (X,\sigma)
\end{equation}
Initially, we will fix the topological type of $\Sigma=\hat\Sigma/\Omega$,
as well as the homology class of any boundary component of $\Sigma$. We 
will always fix the total degree of the map, which in terms of the
graph data is given by $d=\sum d_i$, as well as the total genus of 
$\hat\Sigma$, which is determined by $\hat g =1-k+l+\sum g_i$. The 
target space involution $\sigma$ is determined by \eqref{nonstandard1}, 
\eqref{nonstandard2}, \eqref{nonstandard3}, respectively, and acts only 
on the decoration $\{p_i\}$. The involution on the domain is specified 
by a map on vertices and edges that is compatible with the decoration 
by genus and degree. Moreover, for any fixed edge, we have to specify 
whether the involution acts by $z\to 1/\bar z$ or $z\to  -1/\bar z$ 
on the inhomogeneous coordinate of the corresponding rational curve 
\cite{diaconescu}. The topological type of $\Sigma$ and the map can 
easily be recovered from this data.

Instead of thinking of the equivariant map \eqref{symmetric}, represented
by a real graph. we can also think of a ``half-map'' $\Sigma \to X/\sigma$, 
and accordingly remember only half of the graph, as well as how it is
reflected. This is algorithmically more economical, however one has to 
be extra careful with identifying automorphisms of the graph (see below).

The localization formula then takes the form
\begin{equation}
\eqlabel{sum}
\tilde n^{\Sigma}_d = (-1)^{p(\Sigma)} \sum_{\Gamma} \frac{1}{|\aut\Gamma|}
\int_{\calm_\Gamma}
\frac{\eul(\cale_d)}{\eul(\caln_\Gamma)}
\end{equation}
where the sum is over all real graphs (or half-graphs) of the appropriate type.
Here, $\calm_\Gamma$ is the component of the fixed locus that corresponds to
$\Gamma$, and $\caln_\Gamma$ is the corresponding normal bundle. Depending
on one's point of view, the bundles in \eqref{sum} are either the real bundles 
pulled back via the equivariant map, or the complex bundles pulled back via
the half-maps. The $\eul$'s in \eqref{sum} are the equivariant Euler
classes. We have reserved a sign $(-1)^{p(\Sigma)}$ to be able to adjust
the relative orientation between moduli spaces of maps from different 
worldsheet topologies. This will become important when we sum over the 
worldsheet topologies at fixed Euler characteristic.

To proceed, we will borrow the formulas for the tangent and obstruction 
weights from the cited literature. The signs of the Euler classes coming from
disk components are documented in \cite{psw}, but there will be additional 
signs associated with crosscaps and unoriented loops that we will discuss
below \cite{diaconescu}. An issue that will require renewed attention 
is the appearance of zero weight components when the torus weights are 
specialized from $\TT^n$ to $\TT^{n'}$. One of the results of this
discussion will be that we will never perform integrals over moduli 
spaces of real curves (contracted to a $\sigma$- and $\TT^{n'}$-invariant 
point in target space). The remaining integrals over $\calm_\Gamma$
are performed using Faber's algorithm \cite{faber}. But before we get 
into all these subtleties, and to get oriented about the notation, it will be 
helpful to first discuss a few simple cases. Incidentally, this will 
immediately reveal the fundamental puzzle with the BPS interpretation of 
these invariants, as well as suggest the possible resolution.

One last thing. In the complex case, the localization formula applies
in higher genus only for $X=\text{local $\projective^2$}$. For hypersurfaces,
the formula is only valid in genus $0$, and the methods of
\cite{zinger} have to be invoked in higher genus. In the real case, it
turns out that because of the restriction on the degree of the boundary
components, and the generous treatment of real torus fixed points,
we can actually evaluate {\it certain} classes of higher-genus curves 
using the naive expressions also on the quintic and bicubic.

\begin{figure}[t]
\begin{center}
\epsfig{width=10cm,file=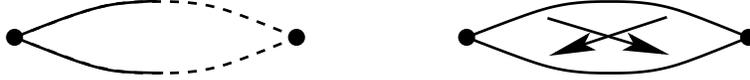}
\caption{Left: The only fixed graph contributing to the annulus invariant 
$\tilde n^{(0,2)}_2$ in degree $2$. Right: The doubled graph can also contribute 
to the Klein bottle invariant $\tilde n^{(1,0)_k}_2$. In the latter case, 
the involution $\Omega$ exchanges the two lines as indicated.}
\label{localization1}
\end{center}
\end{figure}

So consider the degree $2$ invariant of the annulus, with both
boundary components non-trivial in $H_1(L)$. It is easy to see that
(before decoration) there is only type of graph, shown in Fig.\ 
\ref{localization1}. It is straightforward to evaluate the sum \eqref{sum},
and one finds
\begin{equation}
\eqlabel{raw}
\begin{array}{c|c|c|c}
& {\rm quintic} & {\rm bicubic} & \text{local $\projective^2$} \\
\hline
\tilde n^{(0,2)}_2 & -\frac{45}8 & -\frac{9}{8} & \frac{3}{8}
\end{array}
\end{equation}
It is clear that this result is incompatible with the
BPS interpretation expected from \cite{oova,lmv}. One of the features of 
these multi-cover formulas is that the number of boundary components is 
fixed. In particular, no curves of lower genus bubble onto annuli in 
degree $2$. What is worse, the double of the annulus of degree $2$ would
be an elliptic curve of degree $2$. But projective space has no such 
curves! So the (integral) invariant $n^{(1,{\rm real})}_2$ should actually 
vanish.

The most conservative way to reconcile the situation is to realize
that the doubled graph corresponding to our annulus is also equivariant
with respect to a different involution on the worldsheet curve, 
yielding a Klein bottle in the quotient (see right of fig.\ \ref{localization1}).
The weights are equal to those of the annulus, and we identify the signs 
such that the two contributions exactly cancel. At the level of graphs, 
such a cancellation was first described in \cite{diaconescu}, but because 
of a different involution in target space, was interpreted there as a 
cancellation among Klein bottles.

\subsection{Homologically trivial boundaries, and local tadpole cancellation}

After this initial success, let us briefly return to the computations in 
\cite{open,jake,psw}. In these works, disk invariants $\tilde n^{(0,1)}_d$ 
on the quintic were computed for all odd degrees $d\in 2\zet+1$. It was also 
noted that disk invariants of even degree are either ill-defined  because
of mixing with non-orientable worldsheets (crosscaps) or else vanish because
the moduli space is odd-dimensional, and hence any well-defined Euler 
class would be trivial.

\begin{figure}
\begin{center}
\psfrag{2}{$2$}
\epsfig{height=2.5cm,file=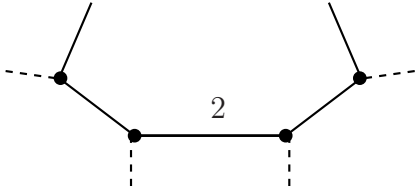}
\caption{A localization graph with a fixed edge of even degree always contributes
equal amounts to a real Gromov-Witten invariant for a worldsheet where $\Omega$ acts 
on that edge with fixed points (a boundary), or without (a crosscap). The signs in 
the localization calculus are such that those contributions exactly cancel.}
\label{localization2}
\end{center}
\end{figure}

Graphically, one can understand this vanishing as follows. Consider
a localization graph with a fixed edge of even degree, equal to $2$ 
in Fig.\ \ref{localization2}. 
Following the localization prescription, the involution on $\hat\Sigma$ can 
act on this fixed edge either by $z\to 1/\bar z$ or $z\to -1/\bar z$, resulting 
in a disk or crosscap component, respectively. We had decided earlier that to 
get an invariant count of real curves, we should combine such contributions.
The best sign is such that they exactly cancel, in agreement with the previous
argument. 

In the computations below, we have extended this prescription 
to any $\Omega$-fixed edge of even degree. The vanishing of the
$\text{(homologically trivial disk)}+\text{crosscap}$ contribution to the 
total invariant can be viewed as the {\it microscopic realization}, graph by 
graph, of {\it tadpole cancellation}, which we will discuss in more detail below. 
Note that our prescription also temporarily addresses the problem that the 
contribution to the equivariant Euler formula from such an even degree edge 
contains (for $n$ odd) a factor $\frac{0}{0}$ and is hence a priori ambiguous. 
\label{forreferee}
We will return to this below.

So let us assume that real graphs with any fixed edge of even degree always 
cancel after summing over different worldsheet involutions. We will also 
assume (and justify later) that graphs with an $\Omega$-invariant
vertex mapping to a $\sigma$- and $\TT^{n'}$-invariant point in 
$\projective^{n-1}$ (those exist for $n$ odd) do not contribute.
As a consequence of this, we compute non-vanishing invariants
$\tilde n^{(g,h)}_d$ by summing over ``orientable half-graphs'' of genus 
$g$ with $h$ ($h>0$) boundary components of odd degree. These correspond 
to real graphs of genus $\hat g=2g+h-1$ that are disconnected when
cut along the $h$ fixed edges. We denote by $\tilde n^{(g,h)_k}_d$ invariants 
obtained from real graphs with $\hat g = 2g+h-1$ that remain connected
after cutting along $h$ fixed edges of odd degree. (In this case, $h$ can
be zero, \cf, eq.\ \eqref{constrained}.) The total degree is
seen to satisfy $d\equiv h\bmod 2$. All other combinations of topological
invariants lead to a vanishing sum over graphs. For future reference we 
record the selection rule
\begin{equation}
\eqlabel{onlyfor}
d\equiv h \equiv \chi \bmod 2
\end{equation}
where we recall that negative $\chi$ is the Euler characteristic of $\Sigma$,
related to the genus of the covering curve by $\chi=\hat g-1$.

\begin{figure}
\begin{center}
\psfrag{2}{$2$}
\psfrag{3}{$3$}
\epsfig{height=4cm,file=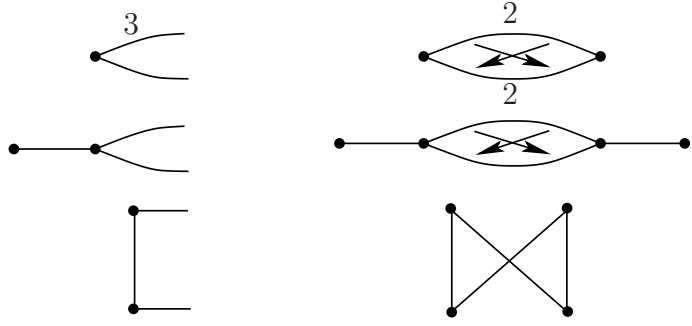}
\caption{Localization graphs contributing to annulus invariant $\tilde n^{(0,2)}_4$
(left) and to the Klein bottle invariant $\tilde n^{(1,0)_k}_4$ (right). In distinction 
to Figs.\ \ref{localization1} and \ref{localization2}, the real graphs are not in 
one-to-one correspondence, and the sum $\tilde n^{(0,2)}_4+\tilde n^{(1,0)_k}_4$ 
does not vanish. (In particular, the graphs on the third line are not identical once the
$\sigma$-symmetric decoration is taken into account.)}
\label{localization3}
\end{center}
\end{figure}

To further clarify these rules, we consider in detail one more example, the
annulus/Klein bottle invariants in degree $4$. There are in each case
three graphs, see Fig.\ \ref{localization3}. Note that there are
even degree edges at the center of the first Klein bottle graph.
However, $\Omega$ acts by exchanging them, so the above vanishing 
rule does not apply. (See next subsection for more details.)
The annulus graphs sum up to
\begin{equation}
\eqlabel{raw2}
\begin{array}{c|c|c|c}
& {\rm quintic} & {\rm bicubic} & \text{local $\projective^2$} \\
\hline
\tilde n^{(0,2)}_4 & -\frac{15525}{16} & -\frac{585}{16} & -\frac{21}{16}
\end{array}
\end{equation}
while the Klein bottles give
\begin{equation}
\eqlabel{raw3}
\begin{array}{c|c|c|c}
& {\rm quintic} & {\rm bicubic} & \text{local $\projective^2$} \\
\hline
\tilde n^{(1,0)_k}_4 & \frac{582725}{16} & \frac{22761}{16} & \frac{117}{16}
\end{array}
\end{equation}
Again, annuli and Klein bottles individually do not make sense from the
point of view of integrality, but their {\it sums}
\begin{equation}
\eqlabel{unity}
2n^{(1,{\rm real})}_4= \tilde n^{(1,0)_k}_4 + \tilde n^{(0,2)}_4 
\end{equation}
are
\begin{equation}
\eqlabel{net}
\begin{array}{c|c|c|c}
& {\rm quintic} & {\rm bicubic} & \text{local $\projective^2$} \\
\hline
2n^{(1,{\rm real})}_4 & 35450  & 1386 & 6
\end{array}
\end{equation}
and integral as advertised.

\subsection{Results}

We will give the explicit localization formula in terms of the 
``half-graphs'' because they are more economical and the signs can
be made more explicit. We will give the weights of the normal
bundle only when there are no collapsed components of higher genus 
(although the total graph might still have non-trivial topology), and 
the weights of $\cale_d$ only for the quintic. The formulas for the 
bicubic are essentially similar. For local $\projective^2$ (and the 
conifold) it also makes sense to evaluate graphs with non-trivial genus
at the vertices. The corresponding formulas can be constructed from,
\eg, \cite{grpa,klza,mayr,diaconescu}.

A graph of requisite type can be thought of as an ordinary graph as in 
say \cite{kontsevich} to which is attached a certain number of 
``half-edges''. Some of these edges represent disks, which requires 
the corresponding degree to be odd. The other half-edges, which can be 
of even or odd degree, are identified pairwise, and represent ``Klein 
edges''. This means that the doubled graph is constructed by reflection 
on the disks and exchange of the pairwise identified half-edges. We can 
count the Klein edges among ordinary edges by remembering that if they 
arise from identification of half-edges attached at vertices decorated 
by $i$ and $j$, then the Klein edge is decorated with $i$ and $\sigma(j)$ 
at its two ends. (This imposes a restriction on the decoration of the 
half-graph.) A similar comment applies to collecting the flags associated 
to the vertices. The contribution of such a graph to the sum \eqref{sum} 
is then given by
\begin{equation}
\eqlabel{formula}
\begin{split}
\int_{\calm_\Gamma} 
\frac{\eul(\cale_d)}{\eul(\caln_\Gamma)}  &= (-1)^{\#{\rm klein}}
\prod_{\rm edges} 
\frac{\displaystyle\prod_{a=0}^{5d} \frac{a\lambda_i + (5d-a)\lambda_j}{d}}
{\displaystyle (-1)^d\frac{(d!)^2}{d^{2d}} (\lambda_i-\lambda_j)^{2d}
\prod_{\topa{k\neq i,j}{a=0}}^d\Bigl(\frac{a}d\lambda_i +
\frac{d-a}d\lambda_j-\lambda_k\Bigr)} 
\\[.2cm] 
\cdot\prod_{\rm disks} &
\frac{\displaystyle\prod_{a=0}^{(5d-1)/2}\frac{a\lambda_{i}+(5d-a)\lambda_{\sigma(i)}}{d}}
{\displaystyle (-1)^{(d-1)/2}\frac{d!}{d^{d}}
(\lambda_{i}-\lambda_{\sigma(i)})^{d}
\prod_{\topa{k\neq i,\sigma(i)}{a=0}}^{(d-1)/2}
\Bigl(\frac ad\lambda_{i} + \frac{d-a}{d}\lambda_{\sigma(i)}
-\lambda_k\Bigr)}
\\
\cdot\prod_{\rm vertices} &
\frac{1}{(5\lambda_v)^{{\rm val}(v)-1}}
 \prod_{j\neq v}(\lambda_v-\lambda_j)^{{\rm val}(v)-1}
\cdot\biggl(\prod_{\rm flags}\frac{d}{\lambda_v-\lambda_j}\biggr)
\biggl(\sum_{\rm flags} \frac{d}{\lambda_v-\lambda_j}\biggr)^{{\rm val}(v)-3}
\end{split}
\end{equation}
The sign at the beginning of \eqref{formula} measures the number of 
Klein edges, which agrees with the rules given in \cite{diaconescu}.
The sign in \eqref{sum} is given by
\begin{equation}
\eqlabel{sign}
(-1)^{p(\Sigma)} = (-1)^{g+\chi-1}
\end{equation}

\begin{figure}
\begin{center}
\psfrag{2}{$2$}
\psfrag{5}{$f=\sigma(f)$}
\psfrag{i}{$i$}
\psfrag{si}{$\sigma(i)$}
\epsfig{width=13cm,file=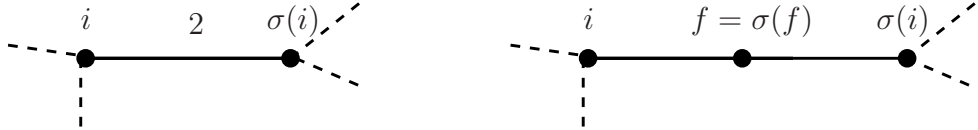}
\caption{Graphs that differ by a local replacement as indicated
actually belong to a one-dimensional fixed locus.}
\label{replacement}
\end{center}
\end{figure}

Before summing these results over all decorated graphs, we need to 
specialize the weights $\lambda_i$, $i=1,\ldots,n$ to those invariant under 
target space involution $\sigma$, \cf, \eqref{nonstandard1}, 
\eqref{nonstandard2}, \eqref{nonstandard3}. For the quintic and local
$\projective^2$, this specialization introduces zero weight components
in the form of $\frac 00$, and we must specify some rules to deal
with this ambiguity. The origin of these unexpected torus-invariant 
directions is the existence of real torus fixed points in 
$\projective^{n-1}$ for odd $n$. The fixed locus thereby acquires
an additional dimension that connects graphs which locally differ
as depicted in Fig.\ \ref{replacement}. (This is the graphical 
representation of the one-parameter family of maps \eqref{onepar}.)
When the edge in question is fixed under the involution, the moduli space
is real one-dimensional, and our rules on tadpole cancellation 
imply a vanishing contribution. When the edge is not fixed under
$\Omega$, our rule is that the complex one-dimensional moduli space 
gives a non-zero contribution that can be taken either from the left 
or the right graph in the figure.

Using this algorithm, we have evaluated the sums \eqref{sum} for a certain 
number of worldsheet topologies and first few non-trivial degrees in each 
case. We summarize the results in the following tables. We also give results 
for the conifold, that has been analyzed previously in \cite{katzliu,grza,mayr}.
In particular, $\nu$ is the choice of torus weights in the fiber of 
$\calo(-1)\oplus\calo(-1)$. Our convention is related to the standard 
framing ambiguity via
\begin{equation}
\nu = 2\nu_{\rm framing} - 1
\end{equation}
There is a symmetry under $\nu\to-\nu$, or $\nu_{\rm framing}\to
1-\nu_{\rm framing}$, hence the results are polynomials
in $\nu^2$.

\renewcommand{\topfraction}{1}

\begin{table}[p]
%\begin{equation}
$$
\begin{array}{|cc||c|c|c|c||}
\hline
 & d & {\rm quintic} & {\rm bicubic} & \text{local $\projective^2$} & {\rm conifold} \\\hline
\tilde n^{(0,2)}_d & 2 & -\frac{45}{8} &- \frac{9}{8}  & \frac 38 & -\frac 18 (1-\nu^2) \\
& 4 & -\frac{15525}{16} &  -\frac{585}{16} &  -\frac{21}{16} &-\frac 1{16} (1-\nu^4)  \\
& 6 & -\frac{6387015}{4} & -\frac{123501}{8}   &  \frac{59}{4} & -\frac{1}{24} (1-\nu^6) \\
& 8 & -\frac{66757119525}{32} & -\frac{150570441}{32}  & -\frac{5781}{32} & 
-\frac 1{32} (1-\nu^8) \\
\hline
\tilde n^{(1,0)_k}_d & 2 & \frac{45}{8} & \frac{9}{8} & -\frac{3}{8} &  \frac 18 (1-\nu^2) \\
& 4 & \frac{582725}{16} & \frac{22761}{16} &  \frac{117}{16} & \frac 1{16} (1-\nu^4) \\
& 6 & \frac{295022375}{4} & \frac{6234093}{8}  & -\frac{411}{4} &  \frac{1}{24} (1-\nu^6)\\
& 8 & \frac{4250971393125}{32} & \frac{10876810761}{32} &  \frac{48981}{32} &
\frac 1{32} (1-\nu^8) \\
\hline\hline
\end{array}
$$
\caption{Localization invariants at $\chi=0$.}
\label{GW1real}
\end{table}
%\end{equation}

%\begin{equation}
\begin{table}[p]
$$
\begin{array}{|cc||c|c|c|c||}
\hline
 & d & {\rm quintic} & {\rm bicubic} &  \text{local $\projective^2$} & {\rm conifold} \\\hline
\tilde n^{(0,3)}_d & 3 & \frac{45}{16} & \frac{3}{16}  &  - \frac{3}{16} &  \frac{1}{48} (1-2\nu^2+\nu^4) \\
& 5 & \frac{5175}8 &  \frac{117}{16}  & \frac{9}8 & \frac{1}{48}(1-3\nu^4+2\nu^6)  \\
& 7 & \frac{36429885}{16} &  \frac{58851}{8}  & -\frac{333}{16} & \frac{1}{48} (1-4\nu^6+3\nu^8) \\
\hline
\tilde n^{(1,1)}_d & 1 & & & \frac 1{12} &  -\frac{1}{12} \\
& 3 & & &  -\frac{3}{8} & -\frac{1}{24}(1+2\nu^2-\nu^4) \\
& 5 &   & & \frac{3}{4} &-\frac{1}{24}(1+3\nu^4-2\nu^6)\\
& 7 & & & \frac{839}{24} & -\frac{1}{24}(1+4\nu^6-3\nu^8) \\
\hline
\tilde n^{(1,1)_k}_d & 3 & -\frac{135}{16} & -\frac{9}{16} & \frac{9}{16} & -\frac{1}{16}(1-2\nu^2+\nu^4) \\
 & 5 & -\frac{866325}{8} & -\frac{22527}{16} & -\frac{171}{8} & -\frac{1}{16}(1-3\nu^4+2\nu^6)\\
& 7 &  -\frac{5109042135}{16} & -\frac{9054153}{8} & \frac{7047}{16} & -\frac{1}{16}(1-4\nu^6+3\nu^8)\\
\hline\hline
\end{array}
$$
\caption{Localization invariants at $\chi=1$.}
\label{GW2real}
\end{table}
%\end{equation}

%\begin{equation}
\begin{table}[p]
$$
\begin{array}{|cc||c|c|c|c||}
\hline
 & d &  {\rm quintic} & {\rm bicubic} & \text{local $\projective^2$} & {\rm conifold} \\\hline
\tilde n^{(0,4)}_d & 4  & -\frac{135}{64} & -\frac{3}{64} & \frac{9}{64} & -\frac{1}{192} (1-\nu^2)^3 \\
& 6 & -\frac{46575}{128} & -\frac{117}{128}  &-\frac{135}{128}   & -\frac{1}{128} (1-\nu^2)^3(1+3\nu^2)\\
& 8 & -\frac{105398415}{32} & -\frac{113931}{32}  & \frac{945}{32} & -\frac{1}{96} (1-\nu^2)^3(1+3\nu^2+6\nu^4)\\
\hline
\tilde n^{(1,2)}_d & 2 & &  & -\frac{5}{128} &  \frac{1}{384}(1-\nu^2)(9-\nu^2) \\
      & 4  & &  & \frac{41}{64} & \frac{1}{192}(1-\nu^2)(4+21\nu^2-9\nu^4)\\
      & 6  & &  & -\frac{81}{128} & \frac{1}{384}(1-\nu^2)(12+12\nu^2+119\nu^4-71\nu^6) \\
\hline
\tilde n^{(1,2)_k}_d & 4 & \frac{405}{32} & \frac{9}{64} & -\frac{27}{32} & \frac{1}{32} (1-\nu^2)^3 \\
& 6 & \frac{15454125}{64} &\frac{66879}{64} &\frac{2997}{64} & \frac{3}{64} (1-\nu^2)^3(1+3\nu^2) \\
\hline
\tilde n^{(2,0)_k}_d & 2 & & & \frac{5}{128}& -\frac{1}{384}(1-\nu^2)(9-\nu^2) \\
& 4  & & & \frac{33}{16} & -\frac{1}{192}(1-\nu^2)(9+11\nu^2-4\nu^4) \\
& 6  & & & -\frac{10953}{64} & -\frac{1}{384}(1-\nu^2)(27+27\nu^2+44\nu^4-26\nu^6)
\\\hline\hline
\end{array}
$$
\caption{Localization invariants at $\chi=2$.}
\label{GW3real}
\end{table}
%\end{equation}

%\begin{equation}
\begin{table}[p]
$$
\begin{array}{|cc||c|c|c|c||}
\hline
 & d & {\rm quintic} & {\rm bicubic}  &\text{local $\projective^2$}& {\rm conifold} \\\hline
\tilde n^{(0,5)}_d & 5 & \frac{2025}{1024} & \frac{15}{1024}& -\frac{135}{1024} & \frac{5}{3072} (1-\nu^2)^4 \\
& 7 & 0 &  -\frac{1911}{5120}  & \frac{1323}{1280}  & \frac{49}{15360} (1-\nu^2)^4(1+4\nu^2) \\
& 9 &  \frac{2510008965}{512} & \frac{2271807}{1280}  & -\frac{10935}{256} & \frac{27}{5120} 
(1-\nu^2)^4(1+4\nu^2+10\nu^4)\\
\hline
\tilde n^{(1,3)}_d & 3 & &  & \frac{3}{128}& -\frac{1}{1152} (1-\nu^2)^2 (11-2\nu^2) \\
 & 5 & &  & -\frac{309}{256}& -\frac{1}{2304} (1-\nu^2)^2 (25+218\nu^2-93\nu^4) \\
\hline
\tilde n^{(2,1)}_d & 1  & & & - \frac{7}{2880} & \frac{7}{2880} \\
& 3 & &  & \frac{79}{2880} & \frac{8+120\nu^2-75\nu^4+10\nu^6}{2880} 
%= \frac{7\cdot 3^2}{2880} - \frac{(1-\nu^2)^2(11-2\nu^2)}{576} 
\\
& 5 & &  & -\frac{59}{128} & \frac{15-48\nu^2+324\nu^4-284\nu^6+63\nu^8}{1152}
\\
\hline
\tilde n^{(1,3)_k}_d & 5 & -\frac{10125}{512} & -\frac{75}{512} & \frac{675}{512} & -\frac{25}{1536} (1-\nu^2)^4 \\
& 7 & -\frac{7816725}{16} & -\frac{360297}{512} & -\frac{11907}{128} & -\frac{49}{1536} (1-\nu^2)^4(1+4\nu^2) \\
\hline
\tilde n^{(2,1)_k}_d & 3 & &  & -\frac{9}{128} & \frac{1}{384}(1-\nu^2)^2(11-2\nu^2) \\
& 5 & &  & -\frac{12723}{1024} & \frac{(1-\nu^2)^2(225+622\nu^2-247\nu^4)}{3072}
\\\hline\hline
\end{array}
$$
\caption{Localization invariants at $\chi=3$.}
\label{GW4real}
\end{table}
%\end{equation}

All these results are compatible with integrality by using the
appropriate multi-cover formula, see section \ref{Bmodel}.
The feature of the conifold results is that the sum over worldsheet
topologies with fixed $\chi$ yields a $\nu$-independent answer
that agrees with the multi-cover formula originally conjectured 
in \cite{oova}. This weight-independence is similar to that noticed 
for freely acting orientifold in \cite{diaconescu}.

%\clearpage

\section{Formal Developments}
\label{formal}

The results of the localization computations of the previous section 
suggest that we study the topological string in the presence of
both D-branes and orientifolds, and give an enumerative or BPS
interpretation only to the total topological string amplitude. In 
our examples, in the A-model, the D-brane and the orientifold plane 
are wrapped on the {\it same Lagrangian}, given as the fixed point set 
of an anti-holomorphic involution. If it makes sense however, wrapping
the same Lagrangian should not be a necessary restriction. For example, 
we know already that disk instantons deform the (superpotential on the) 
Lagrangian viewed as D-brane \cite{open}, whereas the vanishing result 
for holomorphic crosscaps shows that the same Lagrangian viewed as
orientifold plane is not corrected. The natural invariant statement 
is that we should have a similar interpretation of $\text{open}+
\text{unoriented}$ topological string amplitudes whenever the D-brane 
is wrapped in the {\it same homology class} as the orientifold plane.%
\footnote{One can give heuristic arguments why this is sufficient
directly from the point of view of Gromov-Witten theory in the A-model.}

In this section, we will study consequences of this assumption. In 
particular, we will write down holomorphic anomaly equations 
for topological string amplitudes on the general open and non-orientable
worldsheet, extending \cite{bcov2,openbcov}. In the next section, we 
will return to the example and use these holomorphic anomaly equations 
to reproduce and extend the A-model results in the B-model.

\subsection{The tadpole state as a normal function}

We begin the discussion at tree level. Consider the compactification
of the type I string (or type IIB orientifold) on a Calabi-Yau manifold 
$Y$, and recall from \cite{newissues} the formula for the 4-d space-time
superpotential
\begin{equation}
\eqlabel{newissues}
\calw = \int_Y H\wedge \Omega \,,
\end{equation}
where $\Omega$ is the holomorphic three-form and $H$ the RR $3$-form
field strength. As emphasized in \cite{aahv}, this formula is best
viewed as expressing the fact that the superpotential $\calw$ is
``generated by D5-brane charge'', in the following sense. When the 
background contains only D5-branes and O5-planes wrapped on a 
collection of holomorphic curves $C=\sum_i C_i$, tadpole cancellation 
requires that the total homology class vanishes, $[C]=\sum_i [C_i]=0
\in H_2(Y)$. There is then a three-chain $\Gamma$ with boundary $\del
\Gamma=C$, and the formula \eqref{newissues} becomes
\begin{equation}
\eqlabel{normal}
\calw = \int_\Gamma \Omega
\end{equation}
In the presence of O9-planes and D9-branes wrapped on the Calabi-Yau with
some choice of gauge bundle, the formula \eqref{newissues} includes
a contribution from the holomorphic Chern-Simons functional \cite{wittencs},
evaluated at the critical point (\ie, the gauge bundle is holomorphic).
This can again be represented in the form \eqref{normal} for appropriate
choice of $C=\del\Gamma$. Integrals of the form \eqref{normal} are 
related \cite{mowa,openbcov} to what are known mathematically as 
``Poincar\'e normal functions''.

From the holomorphic point of view, the formula \eqref{normal} can be 
understood as arising from a computation of the topological string 
amplitude on the disk and crosscap ($\RP^2$) \cite{sinhav,aahv}. 
This computation is well-defined, and representable geometrically by 
\eqref{normal}, whenever the total topological D-brane charge vanishes. 
Note that in computing the topological charge of the orientifold plane, 
one has to take into account that it fills $d=4$-dimensional space-time. 
This imparts an extra factor of $2^{d/2}=4$ to the O-plane charge.

To make this more tangible, it is convenient to temporarily switch to the 
A-model on the mirror Calabi-Yau $X$. In this case, D-brane charges are 
just homology classes in $H_3(X;\zet)$. The corresponding type IIA
setup contains an O6-plane, which carries $4$ units of D6-brane charge 
(as viewed from the covering space. The gauge group on a tadpole 
canceling D6-brane configuration on top of the O6-plane is $SO(4)$.)

In the previous section, we have seen that to obtain a satisfactory 
integral BPS expansion of the topological amplitudes, computed via
localization in the A-model, we should sum worldsheets of different
topology, at fixed order in string perturbation theory. In forming this 
sum, different number of boundaries simply contribute with a unit 
weight, see \eg, \eqref{unity}. This means that the background contains 
just a single D-brane in the covering space. In other words, for purposes 
of tadpole cancellation in the topological string, {\it the charge of 
the topological orientifold plane is simply equal to its homology 
class}. This value of the topological O-plane charge agrees with that 
identified in \cite{sinhav} by duality with ${\it SO}$/${\it 
Sp}$-Chern-Simons theory.

An elementary way to compute the topological O-plane charge is to use
the parity twisted Witten index of the underlying $\caln=2$ worldsheet
theory. Recall that the Witten index in the open string sector between 
branes $B$ and $B'$ can be computed by an index theorem as an appropriate
inner product of the corresponding D-brane charges \cite{dofi},
\begin{equation}
\eqlabel{wittenindex}
\tr_{\calh_{B,B'}} (-1)^F = \langle \ch(B),\ch(B')\rangle
\end{equation}
Now given an orientifold defined by a parity $P$, the O-plane charge 
$\ch(\text{O-plane})$ is defined by requiring that the
parity twisted Witten index in the open string sector between
any brane $B$ and its parity image $P(B)$ satisfy
\begin{equation}
\eqlabel{paritytwisted}
\tr_{\calh_{B, P(B)}} P (-1)^F = \langle \ch(B),\ch(\text{O-plane})\rangle
\end{equation}
In the A-model, the formulas \eqref{wittenindex} and \eqref{paritytwisted}
reduce simply to the geometric intersection indices between the corresponding
three-cycles, which justifies the above value of the O-plane charge.

To embed this in the superstring, we should consider a situation in 
which the charge of the orientifold plane is equal to that of the corresponding
D-brane. To achieve this, we need to switch from the O6/D6-setup of a type
I/II compactification discussed above, to a situation in which we have
O4-planes/D4-branes wrapped on the 3-cycle, and extended along a 
$2$-dimensional subspace of Minkowski space. Remarkably enough, this 
is {\it exactly} the situation in which we are expecting a BPS interpretation 
of open and unoriented topological string amplitudes \cite{oova,sinhav}.
We have thus completed the circle of observations that began with the A-model 
results in the previous section.

This connection of our findings on tadpole cancellation with the superstring
setup is quite satisfying, but also raises some intriguing questions. First 
of all, it is not immediately clear why the physical setup with O4/D4 requires 
cancellation of RR-tadpoles, because the transverse space is still non-compact. 
To address this, note that from the 4d-perspective, the O/D-string carries 
axionic charge under the appropriate fields from the $\caln=2$ hypermultiplets, 
while the BPS states on the string (that are counted by the topological string 
amplitudes) are charged under the vectormultiplets. Thus the need to cancel the 
tadpoles might indicate that the long range fields of the string do not decay 
fast enough (as there are only two transverse directions) to guarantee decoupling
of vector- and hypermultiplets in the corresponding space-time description.%
\footnote{This possibility was realized in discussions with Juan Maldacena, 
Davide Gaiotto and Andy Neitzke.} 
This is further in agreement with the fact that the hypermultiplet couplings 
are related to the B-model on $X$, and our claim that tadpole cancellation 
amounts to requiring decoupling of A- and B-model.

The reverse puzzle arises if we note that in the physical setup that
actually requires cancellation of tadpoles (namely, with O6/D6), the
normalization of the crosscap is 4 times bigger than the one appropriate
for the topological string. For this, we note that we do not necessarily 
expect integrality of the holomorphic amplitudes from this setup as 
there are no appropriate ``BPS states'' (only domainwalls) that we could 
count. The failure of decoupling of vector- and hypermultiplets is
also not a fundamental problem in the context of $4$-d, $\caln=1$ 
supersymmetry.

In any case, both issues clearly deserve further clarification, which
we will leave for the future. Let us close by summarizing the 
tree-level data from the discussion above and the results on normal 
functions from \cite{mowa,openbcov}. When tadpoles are canceled
in the O4/D4 setup, the $2$-d superpotential on the worldvolume of the
string is computed by the sum of the topological disk and crosscap
amplitude
\begin{equation}
\eqlabel{treelevel}
\calw\equiv \calw_{\rm 2d} \equiv \G {-1}= \frac{1}{\sqrt{2}}\bigl(\F 01 + \R 00\bigr)\,,
\end{equation}
which is mathematically identified as a ``truncated normal function'',
and is the basic holomorphic quantity at tree-level. The normalization
factor $1/\sqrt{2}$ is from eq.\ \eqref{orderchi} and will prove quite 
useful later on. The non-holomorphic data that enters the extended 
holomorphic anomaly equation is the Griffiths infinitesimal invariant 
$\Delta_{ij}$, which is identified physically as the sum of two-point 
functions on the disk plus crosscap. The relation to \eqref{treelevel} 
is
\begin{equation}
\eqlabel{griffiths}
\Delta_{ij} = D_i D_j \calw - C_{ijk} \ee^K G^{k\kb} D_\kb \bar\calw
\end{equation}
where $C_{ijk}$ is the Yukawa coupling (three-point function on the
sphere), and $G_{i\jb}= \del_i\del_\jb K$ the Zamolodchikov special 
K\"ahler metric on the moduli space. The infinitesimal invariant 
satisfies the holomorphic anomaly equation \cite{openbcov}
\begin{equation}
\eqlabel{diskhan}
\del_\ib \Delta_{jk} = - C_{jkl} \Delta_\ib^l \,,
\end{equation}
where $\Delta_\ib^j= \ee^K G^{j \kb} \Delta_{\ib\kb}$. We are now ready for 
loop amplitudes.

\subsection{Holomorphic anomaly at one-loop}

Let us first recall the derivation of the holomorphic anomaly of the
torus amplitude $\Fc 1\equiv \F 10$, see appendix of \cite{bcov1}. This
amplitude is given by a generalized index,
\begin{equation}
\eqlabel{torusamp}
\Fc 1 = \frac 12 \int \frac{d^2\tau}{\tau_2} \tr_{\rm closed} 
\bigl[(-1)^F F_L F_R \ee^{2\pi\ii(\tau L_0-\bar\tau\bar L_0)}\bigr]
\end{equation}
where the integral is over the fundamental domain of the action of
${\it SL}(2,\zet)$ on the upper half-plane, and $\tau_2=\Im \tau$,
and the trace is over the Hilbert space of closed string states.

The torus one-point function is obtained from \eqref{torusamp}
by taking a holomorphic derivative with respect to the closed
string moduli, and can be written as
\begin{equation}
\eqlabel{onepoint}
\del_j\Fc 1 = \frac 12\int \tr_{\rm closed} (-1)^F
\Bigl[ \int \mu G^- \int \bar\mu\bar G^- \phi_j(0) 
 \ee^{2\pi\ii(\tau L_0-\bar\tau\bar L_0)} \Bigr]
\end{equation}
where $\mu,\bar \mu$ are the Beltrami differentials, which are contracted with
$G^-,\bar G^-$ playing the role of the anti-ghosts. Acting with an anti-holomorphic
derivative $\del_\ib$ brings down the BRST-trivial anti-chiral insertion 
$\{G^+,[\bar G^+,\phi_\ib(z)]\}$. By moving the $G^+,\bar G^+$ around the 
trace, this can be converted into the integral of a total derivative, which 
receives a contribution from the boundary of moduli space at $\Im\tau\to\infty$, 
as well as a contact term from the collision of $\phi_j$ and $\phi_\ib$. 
Taken together, the holomorphic anomaly of the torus partition function is
\begin{equation}
\eqlabel{torushan}
\del_\ib\del_j \Fc 1 = \frac 12 \tr C_\ib C_j - \frac{1}{24} 
\tr_{\rm closed} (-1)^F\; G_{\ib j}
\end{equation}
where $C_j$, $C_\ib$ is the representation of the chiral ring on the
RR ground states from the vacuum bundle, see section \ref{background}.

Turning to open/unoriented strings, there are three additional
Riemann surfaces of Euler characteristic $0$: the annulus, the
M\"obius, and the Klein bottle. All three surfaces have a one-dimensional
moduli space of conformal structures, parametrized by $L>0$, and one real 
conformal Killing vector. The three amplitudes are formally written as
\begin{equation}
\eqlabel{amk}
\begin{split}
\annulus \equiv \F 02 &= 
\int_0^\infty \frac{dL}{L} \tr_{\rm open} \bigl[(-1)^F F \ee^{-LH}\bigr] \\
\moebius \equiv \R 01 &= \int_0^\infty \frac{dL}{L} \tr_{\rm open} 
\bigl[(-1)^F F P \ee^{-LH}  \bigr]\\
\klein \equiv \K 10 &= 
\int_0^\infty \frac{dL}{L} \tr_{\rm closed} \bigl[(-1)^F F P \ee^{-LH}  \bigr]
\end{split}
\end{equation}
(For our notation of open/unoriented surfaces, see section \ref{background}.)
In \eqref{amk}, $P$ is the representation of the parity operator on the
space of open/closed string states, and $H=L_0+\bar L_0$ the corresponding 
Hamiltonian.

The holomorphic anomaly equation for these three surfaces can be obtained
by following the same principles as in \cite{bcov1,bcov2}. The moduli
spaces now have two boundaries. The limit $L\to\infty$ corresponds to
factorization in the ``direct'' channel, and $L\to 0$ corresponds to
factorization in the ``transverse'' or ``closed string'' channel, where we 
are using standard textbook terminology.

Factorization of the Klein bottle in the direct channel is essentially
identical to the analysis on the torus, with an additional insertion of
the parity operator in the trace.
\begin{equation}
\eqlabel{kleinhan}
\del_\ib \del_j \klein \underset{{\rm direct}}{\supset}\;
\frac 12 \tr_{\rm closed} \bigl[ C_\ib C_j P \bigr]
\end{equation}
Direct channel factorization of the annulus was shown in \cite{bcov2}
to reduce to the curvature of the $tt^*$-metric in the space of open
string ground states, as a bundle over the closed string moduli space.
Under the claim that only charge $0$ (and $3$) open string ground 
states are relevant, it was argued in \cite{openbcov} that this curvature 
is given by $1/2$ times the closed string result, and hence
\begin{equation}
\eqlabel{anhan}
\del_\ib\del_j\cala\underset{\rm direct}{\supset}
\del_\ib\del_j\tr_{\rm open}\bigl[ (-1)^F \log g_{tt^*}\bigr] 
= \frac N2 G_{\ib j}
\end{equation}
where $N$ is the number of RR ground states of charge $0$, \ie, the
dimension of the gauge group before orientifold. Similarly, direct 
channel factorization of the M\"obius strip takes the form
\begin{equation}
\eqlabel{moehan}
\del_\ib\del_j\calm\underset{\rm direct}{\supset}
\del_\ib\del_j \tr_{\rm open}\bigl[(-1)^F P\log g_{tt^*}\bigr]
= \frac {N^P}2 G_{\ib j}
\end{equation}
where $N^P=N^+-N^-$, and $N^\pm$ is the number of even/odd gauge bosons
under the orientifold. $N^+$ is the dimension of the gauge group after
orientifold.

It is through the transverse channel that we see the appearance of the
potentially harmful tadpoles. In \cite{openbcov}, this was addressed by
restricting attention to the dependence on a discrete open string modulus,
in other words canceling the tadpoles using anti-branes. In practical terms, 
the normal function at tree level is given by the tension of the domainwall 
between two vacua on the brane, instead of the raw superpotential itself. 
The key result is that the contribution to the anomaly can be expressed 
in terms of the infinitesimal invariant as
\begin{equation}
\eqlabel{key}
\del_\ib\del_j \F 02 \underset{\rm transverse}{\supset}
-\Delta_{jk} \ee^K G^{k\kb} \bar \Delta_{\ib \kb}
\end{equation}
In the context of canceling tadpoles with orientifolds, the transverse
channel factorization of the three individual amplitudes might not make
sense any longer. However, following standard considerations, one can see 
that for the sum of the three amplitudes, $\cala+\calm+\calk$,
factorization in the closed string channel can again be expressed as in 
\eqref{key}, where $\Delta_{ij}$ is now the infinitesimal invariant of the 
superpotential \eqref{griffiths}. Taking into account the normalization
convention, we have
\begin{equation}
\del_\ib\del_j\bigl(\cala + \calm + \calk\bigr)
\underset{\rm transverse}{\supset} 
-2 \Delta_{jk} \Delta_{\ib}^k
\end{equation}
Finally, we record the holomorphic anomaly for the total one-loop
amplitude of $\text{open}+\text{closed}+\text{unoriented}$ strings,
\begin{equation}
\G 0 = \frac 12\bigl[\Fc 1 + \cala + \calm + \calk\bigr]
\end{equation}
(Recall our conventions \eqref{total} that the total amplitudes $\G \chi$ 
are indexed by the Euler character of the Riemann surfaces.) We have
\begin{equation}
\begin{split}
\del_\ib\del_j \G 0 =
\frac 14 \tr_{\rm closed} \bigl[C_\ib C_j (1+P)&\bigr] -
 \frac 1{48}  \tr_{\rm closed}  (-1)^F \;  G_{\ib j} \\
+ & \frac 14 \del_\ib\del_j \tr_{\rm open} \bigl[(-1)^F (1+P) \log g_{tt^*}\bigr]
- \Delta_{jk} \Delta_{\ib}^k
\end{split}
\end{equation}

\subsection{The General Holomorphic Anomaly Equation}

Recall that we denote by $\F gh$ the topological string amplitude
on orientable surfaces with $h$ boundary components, and Euler character
$\chi=2g+h-2$, by $\R gh$ the amplitude on non-orientable surfaces with an 
odd number of crosscaps, $h$ boundary components, and Euler character
$\chi=2g+h-1$, and by $\K gh$ the amplitude on non-orientable surfaces
with an even number of crosscaps, $h$ boundary components, and Euler
character $\chi= 2g+h-2$. We now consider $\chi>0$.

The holomorphic anomaly equation for $\Fc g\equiv \F g0$ is given by
\cite{bcov2}
\begin{equation}
\del_\ib \Fc g = \frac 12 \sum_{g_1+g_2=g}
 C_\ib^{jk} \Fc {g_1}_j \Fc {g_2}_k +
\frac 12 C_\ib^{jk} \Fc {g-1}_{jk} 
\end{equation}
The first term originates from the closed string degeneration in
which the Riemann surface splits into two components, of genus $g_1$
and $g_2$ ($g_i>0$), while the second comes from the pinching of
a handle that reduces the genus by $1$.

This equation was extended in \cite{openbcov} to orientable Riemann
surfaces with $h>0$, with tree-level data given by the tension of a
domainwall between two brane vacua. The extension reads
\begin{equation}
\eqlabel{extension}
\del_\ib \F gh = 
\frac 12 \sum_{
\topa{\topa{g_1+g_2=g}{h_1+h_2=h}}{2g_i+h_i>1}} 
C_{\ib}^{jk} \F {g_1}{h_1}_j \F {g_2}{h_2}_k +
\frac 12 C_{\ib}^{jk} \F {g-1}h_{jk} - 
\Delta_{\ib}^j \F g{h-1}_j  \,,
\end{equation}
Again, the first two terms come from closed string degenerations,
while the last comes from the shrinking of a boundary component
to zero size. Degenerations in the open string channel were argued
in \cite{openbcov} to not contribute generically.

It is not hard to see what must be the extension of these results to
the type of orientifold background that we discussed above, with tree-level 
data given by \eqref{griffiths}. It suffices to understand how the various
non-orientable worldsheets or their symmetric covers can degenerate. Again,
we neglect degenerations in the open string channel.

Under a closed string degeneration (growth of an infinitely long tube) a
non-orientable Riemann surface with an odd number of crosscaps $\Sigma^{(g,h)_r}$
can split into two components, at least one of which must be non-orientable.
Or a handle can pinch, reducing the genus by $1$. In the latter case,
the pinching handle can be straight, or it can be a Klein handle (parity
reversing). The remaining Riemann surface is of type $(g-1,h)_r$ in 
both cases. Let us introduce the notation for the parity-twisted Yukawa 
coupling from eq.\ \eqref{convenient},
\begin{equation}
\eqlabel{paritytwist}
B_{ijk} = C_{ijl} P^l_k
\end{equation}
as well as its cousins with raised indices. We can then write the corresponding
contribution to the holomorphic anomaly of the amplitude $\R gh$ as
\begin{equation}
\eqlabel{Rhan}
\begin{split}
\del_\ib \R gh \underset{\rm closed}{\supset}
\sum_{\topa{g_1+g_2=g}{h_1+h_2=h}}
C_{\ib}^{jk} \K {g_1}{h_1}_j \R {g_2}{h_2}_k +
& \sum_{\topa{g_1+g_2=g}{h_1+h_2=h}}
C_{\ib}^{jk}  \F {g_1}{h_1}_j  \R {g_2}{h_2}_k \\
&\qquad + \frac 12 C_{\ib}^{jk} \R {g-1}{h}_{jk}
+ \frac 12 B_{\ib}^{jk} \R {g-1}h_{jk}
\end{split}
\end{equation}
Non-orientable Riemann surfaces with an even number of crosscaps, 
$\Sigma^{(g,h)_k}$, have several more possible types of closed string
degenerations, and we obtain
\begin{equation}
\eqlabel{Khan}
\begin{split}
 & \del_\ib  \K gh \underset{{\rm closed}}{\supset}
\sum_{\topa{g_1+g_2=g}{h_1+h_2=h}} C_{\ib}^{jk}
\K {g_1}{h_1}_j \F {g_2}{h_2}_k +
\frac 12 \sum_{\topa{g_1+g_2=g-1}{h_1+h_2=h}} C_{\ib}^{jk}
\R {g_1}{h_1}_j \R {g_2}{h_2}_k  \\ 
&  +  \frac 12  \sum_{\topa{g_1+g_2=g}{h_1+h_2=h}} C_{\ib}^{jk}
\K {g_1}{h_1}_j \K {g_2}{h_2}_k    
 + \frac 12 C_{\ib}^{jk} \K {g-1}h_{jk} +
\frac 12 B_{\ib}^{jk} \K {g-1}h_{jk} +
\frac 12 B_\ib^{jk} \F {g-1}h_{jk} 
\end{split}
\end{equation}

\begin{figure}[t]
\begin{center}
\epsfig{width=13cm,file=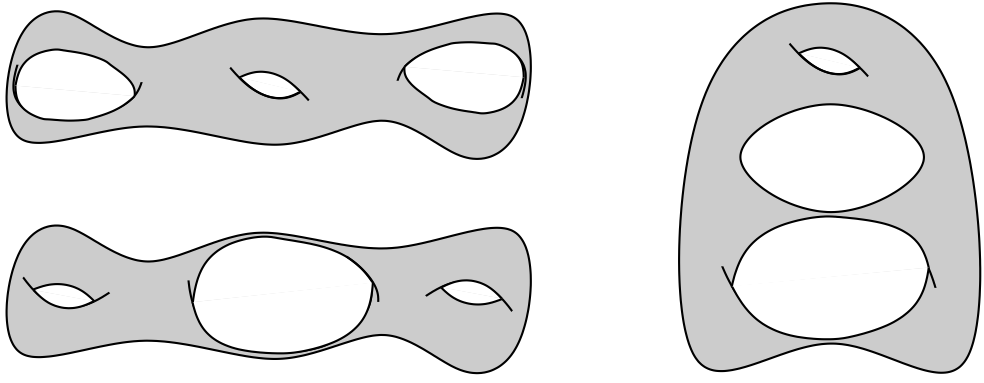}
\caption{Three degenerations of a Klein surface $\Sigma$ of type $(2,0)_k$, 
viewed from the covering symmetric Riemann surface, of genus $\hat g=3$. 
Upper left: $\Sigma$ degenerates to a Klein bottle via pinching of a handle. 
Lower left: $\Sigma$ degenerates to a torus via pinching of a Klein handle. Right:
$\Sigma$ degenerates to a Klein bottle via pinching of a Klein handle.}
\label{pinchings}
\end{center}
\end{figure}

To clarify that the pinching of a Klein handle is a different limit than
the pinching of a straight handle, we show the corresponding degenerations 
of the covering symmetric Riemann surface in the case $\hat g = 3$ in Fig.\ 
\ref{pinchings}.

Note that unstable ``tadpole'' degenerations involving single disks or 
crosscaps with just one closed string insertion are excluded from the above
formulas. To make sense of these degenerations, we note that the corresponding 
singular Riemann surfaces always arise as the {\it common limit} of two 
worldsheets of {\it different topology}. Namely, a real node of the covering
surface can be smoothed to yield either a disk or a crosscap in the quotient.
This is a principle that we have encountered in our A-model discussions in
section \ref{Amodel}. By adding the two contributions, we obtain the insertion 
of a tadpole state on the limiting Riemann surface. Explicitly,
\begin{equation}
\eqlabel{tadpoles}
\begin{split}
\del_\ib\bigl(\F gh + \R g{h-1}\bigr) & 
\underset{\rm tadpole}{\supset} -\sqrt 2\Delta_\ib^j \F g{h-1}_j \\
\del_\ib\bigl( \K gh + \R g{h-1}\bigr) & 
\underset{\rm tadpole}{\supset} -\sqrt 2\Delta_\ib^j \K g{h-1}_j \\
\del_\ib\bigl( \K gh + \R {g-1}{h+1}\bigr) & 
\underset{\rm tadpole}{\supset} -\sqrt 2\Delta_\ib^j \R {g-1}h_j
\end{split}
\end{equation}
where the $\sqrt{2}$ again comes from the normalization of the superpotential
\eqref{treelevel}.

Let us now assemble these various pieces and consider the holomorphic
anomaly for the total topological string amplitudes at order $\chi>0$ 
in string perturbation theory. As discussed in section \ref{background},
these are given by
\begin{equation}
\eqlabel{normalization}
\G \chi = \frac{1}{2^{\frac\chi 2+1}}
\Bigl[\calf^{(g_\chi)} 
+ \sum_{2g+h-2=\chi} \F gh 
+ \sum_{2g+h-1=\chi} \R gh
+ \sum_{2g+h-2=\chi} \K gh \Bigr]
\end{equation}
By combining the formulas \eqref{extension}, \eqref{Rhan}, \eqref{Khan},
and \eqref{tadpoles}, we obtain 
\begin{equation}
\eqlabel{totalhan}
\del_\ib \G\chi = 
\frac 12 \sum_{\chi_1+\chi_2=\chi-2}
C_\ib^{jk}\G{\chi_1}_{j}\G {\chi_2}_k
+\frac 14 \bigl(C_\ib^{jk}+B_\ib^{jk}\bigr) \G {\chi-2}_{jk}
- \Delta_\ib^j \G {\chi-1}_j
\end{equation}

It is a pleasant surprise that this final form of the holomorphic anomaly 
equation is very similar to the extended holomorphic anomaly of \cite{openbcov}, 
see \eqref{extension}. Note that the prefactor in the second term
\begin{equation}
\eqlabel{prefactor}
{C^P}_\ib^{jk} = \frac 12 \bigl(C_\ib^{jk}+B_\ib^{jk}\bigr) 
= C_\ib^{jl}\; \frac{\delta_l^k + P_l^k}{2}
\end{equation}
is simply the projection of the Yukawa coupling onto the parity-invariant 
states. Moreover, we note that we may also endow $C_\ib^{jk}$ in the first
term of \eqref{totalhan} with the same projector \eqref{prefactor}. This is
because the one-point functions $\G \chi_j$ with insertion of a parity-odd 
field must vanish identically. (The two-point function $\G\chi_{jk}$ might not 
vanish when both fields are odd, so here we must use that the projector comes 
out of the degeneration of Riemann surfaces.) Finally, we can also insert
a projector in front of the last term in \eqref{totalhan}, 
\begin{equation}
{\Delta^P}_\ib^j = \Delta_\ib^l \; \frac{\delta_l^j + P_l^j}{2}
\end{equation}
to obtain
\begin{equation}
\eqlabel{identical}
\del_\ib \G\chi = 
\frac 12 \sum_{\chi_1+\chi_2=\chi-2}
{C^P}_\ib^{jk}\G{\chi_1}_{j}\G {\chi_2}_k
+\frac 12 {C^P}_\ib^{jk} \G {\chi-2}_{jk} - {\Delta^P}_\ib^j \G {\chi-1}_j
\end{equation}
This holomorphic anomaly equation is even closer to \eqref{extension},
with the important difference that fields that are projected out by the 
orientifold have completely decoupled, as we should have expected. 

As is well-known, parity defines a holomorphic involution
\begin{equation}
P:M\to M
\end{equation}
of the moduli space of the topological string, and the invariant subspace,
\begin{equation}
M^P = \{P(m)= m\} \subset M
\end{equation}
is the moduli space of the orientifold.

As a consequence of these observations, all the results on solving the 
holomorphic anomaly equation by Feynman diagrams \cite{bcov2,openbcov,
coy,newa}, and on the polynomial structure of the solutions \cite{yayau,
alim,konishi} will carry over with no essential modification to the 
orientifold situation. In particular, since $M^P$ is a special K\"ahler 
submanifold of the special K\"ahler manifold $M$, the special geometry
relation
\begin{equation}
\eqlabel{specgeom}
{{R_{\ib j}}^k}_l = C_{\ib}^{mk} C_{jml} - G_{\ib j} \delta^k_l - G_{\ib l} 
\delta^k_j
\end{equation}
will continue to hold on the orientifold moduli space, and allow for
the construction of propagators, terminators, \etc. In the examples 
below, we will however only deal with one-parameter models, with $M^P=M$, 
so we would have little use for developing this general formalism 
explicitly.

A notable difference to the works \cite{coy,newa} is that \eqref{identical}
is an equation only for the total topological amplitude \eqref{normalization},
and not for the $\F gh$, \etc, individually. Namely, tadpole cancellation
does not allow introducing a free 't Hooft parameter into \eqref{total},
in addition to the string coupling. Nevertheless, we can still compute 
individual amplitudes $\F gh$ for fixed $h$ if we restrict to dependence 
on the discrete open string moduli, as was done in \cite{openbcov}. This 
can be seen as a replacement for inserting continuous open string moduli 
on the worldsheet boundaries, which we have argued before generically%
\footnote{Of course, when there {\it are} non-trivial open string moduli 
present, we can insert those, too.} decouple from the topological 
amplitudes.

\section{The Examples in the B-model}
\label{Bmodel}

The topological B-model is governed at closed string tree-level by 
special geometry, which coincides for Calabi-Yau threefolds with 
the mathematical theory of variation of Hodge structure \cite{dave}. 
The workhorse \cite{cdgp} is the Picard-Fuchs differential equation 
satisfied by the periods of the holomorphic three-form. By extension, 
the open string tree-level information (domainwall tensions) can also 
be obtained by solving an appropriate differential equation \cite{lmw}, 
which in the absence of open string moduli is simply an inhomogeneous 
version of the Picard-Fuchs equation \cite{open,mowa}. This might be 
referred to as $\caln=1$ special geometry, and is related to the
mathematical theory of Poincar\'e normal functions.

The extension to orientifold backgrounds is rather straightforward. As 
explained in section \ref{formal}, the tree level data now consists of
the full superpotential, and when tadpoles are canceled still fits 
into the framework of normal functions.

Loop amplitudes can be computed by solving the holomorphic anomaly
equations, for which there are several well-known techniques. The 
holomorphic ambiguities can be fixed by imposing appropriate boundary 
conditions at the various singular loci in moduli space. One of the 
outcomes of our computations in this section is that the holomorphic 
ambiguities for open and unoriented string amplitudes appear to be often 
simpler than their closed string counterparts. 

\subsection{Tree-level data}

Recall that our three examples were defined in the A-model as
the quintic in $\projective^4$, the bicubic in $\projective^5$,
and the total bundle of the canonical bundle over $\projective^2$ 
(local $\projective^2$). The involution defining the orientifold
came in each case from the standard complex conjugation of the 
corresponding projective space. The D-branes are wrapped on the 
fixed locus of this involution, and to cancel the tadpoles we need 
exactly one D-brane in the covering space. In each case, there
are two brane vacua, and the topological string amplitudes
depend on the discrete parameter $\epsilon=\pm 1$ in addition to
the bulk K\"ahler parameter $t$.

The mirror of the quintic is the mirror quintic, which can be obtained 
by blowing up singularities of an appropriate orbifold of a one-parameter
family of quintics. The Picard-Fuchs operator is
\begin{equation}
\call = \theta^4 - 5z (5\theta+1)(5\theta+2)(5\theta+3)(5\theta+4)
\end{equation}
Where $\theta=z d/dz$, and $z$ is the complex structure parameter of
the mirror family, which is related to the K\"ahler parameter of the
quintic by the mirror map,
\begin{equation}
\eqlabel{mirrormap}
t(z) = \frac{\varpi_1(z)}{\varpi_0(z)}
\end{equation}
Here $\varpi_0(z)$ and $\varpi_1(z)$ are the analytic and first
logarithmic solutions, respectively, of the Picard-Fuchs differential 
equation
\begin{equation}
\call \varpi(z) = 0
\end{equation}
around $z=0$.

The mirror of the two brane vacua on the real quintic is a certain pair 
of matrix factorizations of the Landau-Ginzburg superpotential 
\cite{strings}. The corresponding normal function is studied in 
detail in \cite{mowa}. It can be represented as the difference of 
curves $[C_+-C_-]\in{\rm CH}^2(Y)$, where
\begin{equation}
\begin{split}
C_\pm = \{x_1+x_2=0,\;  & x_3+ x_4= 0,x_5^2\pm\sqrt{5\psi} x_1x_3 = 0 \} \\
& \subset Y = 
\{ x_1^5+x_2^5+x_3^5+x_4^5+x_5^5 - 5\psi x_1x_2x_3x_4x_5 = 0\}
\end{split}
\end{equation} 
($z=(5\psi)^{-5}$). The domainwall tension $\calt=\int_\Gamma\Omega$
(with $\del\Gamma= C_+-C_-$) satisfies the inhomogeneous Picard-Fuchs 
equation
\begin{equation}
\eqlabel{inhpf}
\call \calt(z) = c \sqrt{z}
\end{equation}
where $c=\frac{15}{16\pi^2}$. 

The mirror of the orientifold action on $X$ is just the trivial involution on 
$Y$ \cite{brho}, acting on D-branes by duality. In other words, the superstring
version would simply be the type I string compactified on $Y$. This means
that topologically, the O-plane charge can be expressed in terms of
the tangent bundle of the mirror quintic 
\begin{equation}
\eqlabel{topcharge}
\ch(\text{O-plane}) \propto \sqrt{L( \textstyle{ \frac 14} TY)}
\end{equation}
As we have mentioned in section \ref{background}, it is not clear in general
how the orientifold plane is represented {\it holomorphically}. However,
in the context of type I on the quintic, we have the more elementary 
expression \eqref{newissues} for the superpotential \cite{newissues}, which 
can be reduced to the statement that \eqref{topcharge} is also valid holomorphically. 
Since by the adjunction formula, the Chern classes of the quintic come from 
projective space, they are independent of the complex structure parameter $z$. 
(Although we have not checked this explicitly, we expect that
the orbifold by $(\zet_5)^3$ will not affect this conclusion, and at most
contribute an overall normalization factor, rather as in \cite{mowa}.)
As a consequence, the superpotential $\calw$, see eq.\ \eqref{treelevel}, 
satisfies the same differential equation \eqref{inhpf}, with $c\to \ii c/2$, 
where we have inserted a factor of $\ii$ for consistency with previous work,
and we have removed the factor of $\sqrt{2}$ from $\calw$ because this is
more convenient when working with the normalization \eqref{found} for
the topological string amplitudes. As normalization benchmark, we give here 
the expansion of the normalized Yukawa coupling and normalized infinitesimal 
invariant in terms of $q=\exp(2\pi\ii t)$
\begin{equation}
\begin{split}
C_{ttt} & = 5 + 2875q + 4876875 q^2 + \cdots \\
-\ii \Delta_{tt} &= \frac{15}2 q^{1/2} + 3450 q^{3/2} + 6801570 q^{5/2} + \cdots
\end{split}
\end{equation}
and turn to the other examples.

The mirror of the bicubic was studied in \cite{lite}. The Picard-Fuchs
operator is
\begin{equation}
\call = \theta^4 - 9z (3\theta+1)^2 (3\theta+2)^2
\end{equation}
The mirror of the real bicubic as orientifold and D-brane has not been 
studied in detail yet. It should not be hard to obtain the explicit 
representatives of the normal function, however we will not really need 
this for the computational purposes in this section. It suffices to note 
that the inhomogeneous Picard-Fuchs equation governing the tree level 
data is
\begin{equation}
\call \calw (z) = \frac{9\ii}{32\pi^2}\sqrt{z}
\end{equation}
The normalization factor can be checked by computing the first term
in the Gromov-Witten expansion of the (normalized) superpotential
\begin{equation}
\eqlabel{shende}
-4\pi^2 \ii \frac{\calw (z(q))}{\varpi_0(z)} =
\sum_{d\;{\rm odd}} \tilde n^{(0,1)}_d q^{d/2}
= 2 \sum_{\topa{d\;{\rm odd}}{k\; {\rm odd}}} \frac{1}{k^2} n^{(0,{\rm real})}_d 
q^{d k/2}
\end{equation}
in the A-model. In fact, it has been shown \cite{shende} that the theorems of 
\cite{psw} also hold for the bicubic, \ie, \eqref{shende} is valid rigorously 
to all orders. The second step in \eqref{shende} is the BPS expansion and 
the $n^{(0,{\rm real})}_d$ are (conjecturally) all integer. 

The mirror of local $\projective^2$ is captured, see \eg, \cite{klza}, by the 
family of elliptic curves with equation
\begin{equation}
\eqlabel{family}
x_1^3+x_2^3+x_3^3 - 3\psi x_1x_2x_3 = 0
\end{equation}
where the $x_i$ are viewed as $\complex^*$-variables. The Picard-Fuchs operator 
is
\begin{equation}
\call = \theta^3 - 3z\theta(3\theta+1)(3\theta+2) 
\end{equation}
(Where, similarly to as before $z=(3\psi)^{-3}$, and $\theta=zd/dz$.) To obtain 
the inhomogeneous version, we can take the same shortcut as for the bicubic. 
The extension reads
\begin{equation}
\call \calw(z) = -\frac{\ii}{16\pi^2} \sqrt{z}
\end{equation}
In fact, it is not hard to check that the corresponding normal function can
be represented by the two points on the Riemann surface \eqref{family}
\begin{equation}
p_\pm = \{ x_1+x_2=0, x_3 = \pm\sqrt{3\psi} x_1\} \subset 
\{ x_1^3+x_2^3+x_3^3 - 3\psi x_1x_2x_3 = 0 \}
\end{equation}
Namely, we can write the domainwall tension between brane vacua as
\begin{equation}
\calt = \int_{p_+}^{p_-} \lambda
\end{equation}
where $\lambda \propto \log(x_2/x_3)dx_1/x_1$ is the reduction of the 
holomorphic three-form to the curve.

An alternative way to obtain the inhomogeneous term in the Picard-Fuchs
equation is to study carefully the monodromy properties of the domainwall
tension/superpotential as an analytic function over the entire (thrice-punctured)
$z$-plane. This was done for the quintic in \cite{open}. The order-two 
branch points at $z=0$ and $z=\infty$ are easily identified, and the
prefactor $c$ follows from requiring integrality of the monodromy matrices
around the conifold. This exercise could be repeated for the bicubic and 
local $\projective^2$, but we will omit this here.

We conclude this subsection by giving the explicit results for the
genus $0$ real enumerative invariants for the three models discussed
above in table \ref{zeroquintic}.

\begin{table}[t]
\begin{center}
\begin{tabular}{|c|r|r|r|}
\hline
 $d$ & quintic  &   bicubic & local $\projective^2$ \\\hline
1   & 15                              & 9               & $-1$ \\
3   & 765                             & 90              & $1$ \\
5   & 544125                          & 15759           & $-5$ \\
7   & 487998390                       & 3297987         & $42$ \\
9   & 536543881350                    & 841201389       & $-429$ \\
11  & 664513551962205                 & 241496789706    & $4939$ \\
\hline
\end{tabular}
\caption{Genus $0$ real BPS invariants $n^{(0,{\rm real})}_d$}
\label{zeroquintic}
\end{center}
\end{table}

\subsection{One-loop}

The general solution of the holomorphic anomaly of the torus partition function 
\eqref{torushan} is
\begin{equation}
\eqlabel{torussol}
\Fc 1 = \frac 12 \log\Bigl[ \det G_{\ib j}^{-1} \ee^{K(3+n-\frac 1{12}\chi)} |f|^2\Bigr]
\end{equation}
where $G_{\ib j}$ is the special K\"ahler metric on moduli space with K\"ahler
potential $K$. The holomorphic ambiguity $f$ can be fixed by imposing the
appropriate behavior at the boundaries of moduli space. Of relevance for
the one-parameter models are large volume, conifold and orbifold point. 
In the holomorphic limit, and with above conventions, one obtains
\begin{equation}
\Fc 1 \underset{\rm hol.}{\to} \;\; \frac 12 \log 
\Bigl[\Bigl(\frac qz\frac{dz}{dq}\Bigr)\,
(\varpi_0)^{\frac\chi{12}-4} z^{-c_2/12} {\rm diss}^{-1/6}
\Bigr]
\end{equation}
where $c_2$ is the second Chern class of the model, and
${\rm diss} = (1-5^5 z)$, $(1-3^6z)$, and $(1-3^3 z)$ for quintic,
bicubic, and local $\projective^2$, respectively. 

We now turn to the open/unoriented amplitudes at one-loop. As we have
advertised before, only the total amplitude 
\begin{equation}
2\G 0 - \Fc 1 = \cala + \calm + \calk
\end{equation}
($\text{annulus}+\text{M\"obius}+\text{Klein bottle}$) will admit an integral 
expansion in the sense conjectured in \cite{oova}. However, as we have also 
emphasized, there are also certain individual parts of the amplitude that 
make sense and can be computed separately. Specifically, by considering
pairs of branes/antibranes with discrete Wilson lines $\epsilon_1,
\epsilon_2$, it makes sense to isolate a term that depends on 
these discrete parameters by
\begin{equation}
\eqlabel{exact}
\G 0(\epsilon_1,\epsilon_2) -
\G 0(-\epsilon_1,\epsilon_2) - \G 0(\epsilon_1,-\epsilon_2)
+ \G 0 (-\epsilon_1,-\epsilon_2) =  4 \F 02 \epsilon_1 \epsilon_2
\end{equation}
which is essentially how we define $\F 02$ in our examples.

At the second stage, we compute the amplitude for the Klein bottle, 
$\calk\equiv \K 10$. (Or more precisely, the amplitude for the 
Klein bottle plus the $\epsilon$-independent part of the annulus and 
M\"obius amplitude. Note that the M\"obius strip does not make any 
additional contribution in the transverse channel because the tadpole 
state (or infinitesimal invariant) has no $\epsilon$-independent part.) 
Finally, we construct the total topological amplitude of our orientifold 
with one background D-brane. This is of the form
\begin{equation}
2 \G 0 (\epsilon) - \Fc 1 = \calk + \F 02 \epsilon^2 = \calk + \F 02
\end{equation}
with no apparent $\epsilon$-dependence.

The solution of the holomorphic anomaly equation of the annulus is 
for the one-parameter models, and in the holomorphic limit,
\begin{equation}
\eqlabel{annsol}
\cala_t = \F 02_t =  \frac 12 \Delta_{tt}^2 C_{ttt}^{-1} + f^{(0,2)}_t
\end{equation}
where $f^{(0,2)}$ is a holomorphic ambiguity. In \cite{openbcov}, it was 
originally claimed that there is an additional term on the RHS. With this 
additional term, and with a naive ansatz for the BPS expansion of $\F 02$, 
the holomorphic ambiguity $f^{(0,2)}$ could be fixed such that all expansion 
coefficients were integer. However, the situation considered in \cite{openbcov} 
was that of \eqref{exact}, so the effective dimension of the gauge group should
actually have been zero. The same statement holds in the orientifold setup 
with exactly one D-brane in the covering space (\ie, $N^+=0$ in \eqref{moehan}).

In all three examples that we study, we find that the low degree Gromov-Witten 
invariants $\tilde n^{(0,2)}_d$ of table \ref{GW1real} are reproduced by 
\eqref{annsol} with $f^{(0,2)}\equiv 0$. This pattern persists for all 
genus $0$ amplitudes with an {\it arbitrary} number of boundaries,
to which we will return below. The precise way is
\begin{equation}
\F 02 = \sum_{d\;{\rm even}} \tilde n^{(0,2)}_d q^{d/2}
\end{equation}

The Klein bottle contribution \eqref{kleinhan} to the holomorphic anomaly 
of $\G 0$ can be integrated by the same procedure as that leading to 
\eqref{torussol}. We first note that when the orientifold projection acts 
trivially on the moduli space, we have
\begin{equation}
\eqlabel{kleinzwei}
\del_\ib\del_j \klein = \frac 12 \tr \bigl[C_\ib C_j P\bigr] =
\frac 12 C_{jkl} C_\ib^{kl} - G_{\ib j}
\end{equation}
This can be seen by using that with respect to the convenient $tt^*$-basis 
for the Ramond-Ramond ground states, the chiral ring multiplication matrices 
take the form
\begin{equation}
\eqlabel{struccon}
C_i = \begin{pmatrix}
0&0&0&0\\
\delta_i^l & 0 & 0 & 0\\
0 & C_{im}^{\;\;\;\;\lb} & 0 & 0\\
0&0&G_{i\mb}&0
\end{pmatrix}\,,
\qquad
C_\jb =
\begin{pmatrix}
0&G_{\jb m}&0&0\\
0&0&C_{\jb\mb}^{\;\;\;\;l} &0\\
0&0&0&\delta_{\jb}^\lb\\
0&0&0&0
\end{pmatrix}
\end{equation}
and $P$ is represented by \eqref{ofform}. In turn, \eqref{kleinzwei} can be 
integrated by using the special geometry relation for the Ricci tensor, 
\eqref{specgeom}. This yields
\begin{equation}
\calk = \frac 12 \log \bigl( \det G_{\ib j}^{-1} \ee^{K(n-1)} |f^{(1,0)_k}|^2\bigr)
\end{equation}
We can fix the holomorphic ambiguity of the Klein bottle by expanding
$\calk$ around $q=0$ in the holomorphic limit and compare with the
localization results of section \ref{Amodel}.
\begin{equation}
\calk \underset{\rm hol.}{\to} \sum_{d\; {\rm even}} \tilde n^{(1,0)_k}_d q^{d/2}
\end{equation}
It turns out that in all three cases, the holomorphic limit can be written
as
\begin{equation}
\calk \to - \frac 12 \log \Bigl[\Bigl(\frac qz\frac{d z}{dq}\Bigr) {\rm diss}^{-1/4}
\Bigr]
\end{equation}
where ${\rm diss}=0$ describes the conifold locus in moduli space (see
above). We see that as for the torus amplitude, the Klein bottle exhibits
a universal singular behavior associated with the conifold. It will not
be hard, but crucial for future developments, to pinpoint the microscopic 
origin of this universality.

Finally, we sum annulus and Klein bottle and expand in the holomorphic limit
\begin{equation}
 \cala + \calk = 2 \sum_{\topa{d\;{\rm even}}{k\;{\rm odd}}}
\frac 1k n^{(1,{\rm real})}_d q^{dk/2}
\end{equation}
to extract the real BPS invariants $n^{(1,{\rm real})}_d$. We have checked 
integrality up to $d\sim 50$, and list the first few in table \ref{genusone}.
\begin{table}[t]
\begin{center}
\begin{tabular}{|c|r|r|r|}
\hline
 $d$ & quintic  &   bicubic & local $\projective^2$ \\\hline
2   &  $0$                      & 0                   &  0 \\
4   &  $17725$                  & $693$               &  $3$ \\
6   &  $36079420$               & $381912$            &  $-44$ \\
8   &  $65378348025$            & $167597505$         & $675$ \\
10  &  $116755627418596$        & $70912518192$       & $-10596$ \\
12  &  $209184366237053675$     & $29843206833573$    & $169815$\\
\hline
\end{tabular}
\caption{Genus $1$ real BPS invariants $n^{(1,{\rm real})}_d$}
\label{genusone}
\end{center}
\end{table}

\subsection{Next loop}

At the next order in perturbation theory, $\chi=1$, there are three non-trivial 
worldsheets contributing to the total amplitude. (In our examples, all amplitudes 
$\R gh$ with an odd number of crosscaps vanish. This follows from the vanishing of
the crosscap contribution to the superpotential, discussed around eq.\ \eqref{topcharge},
together with the recursive nature of the holomorphic anomaly equations. In the 
A-model, the vanishing of the $\R gh$ is a consequence of local tadpole cancellation, 
as discussed on page \pageref{forreferee}.) Consider
\begin{equation}
2 \G 1 = \F 03 + \F 11 + \K 11
\end{equation}
The solution of the holomorphic anomaly equation for $\F 03$ is
\begin{equation}
\eqlabel{turnsout}
\begin{split}
\F 03 &= - \F 02_j\Delta^j - \frac 12\Delta_{jk}\Delta^j\Delta^k
-\frac 16 C_{jkl}\Delta^j\Delta^k\Delta^l + {\it hol.\ amb.} \\
&=  -\F 02_t\Delta^t - \frac 13\Delta_{tt}\Delta^t\Delta^t
\end{split}
\end{equation}
where in the second line we have specialized to the one-parameter models.
Again, it turns out that the localization results are reproduced exactly
by \eqref{turnsout} with vanishing holomorphic ambiguity $f^{(0,3)}=0$.

The oriented one-loop amplitude with one boundary is given by
\begin{equation}
\begin{split}
\F 11 &= \frac 12 S^{jk}\Delta_{jk} - \F 10_j\Delta^j
+\frac 12 C_{jkl} S^{kl} \Delta^j -\Bigl({\frac{\chi}{24}}-1\Bigr)\Delta 
+ {\it hol. amb.}\\
&=  -\F 10_t\Delta^t -\Bigl(\frac{\chi}{24}-1\Bigr) \Delta
+ f^{(1,1)}
\end{split}
\end{equation}
(As usual, the term with $\chi/24$ is absent for local $\projective^2$ for 
appropriate choice of terminator.) For the quintic and bicubic, we have
no localization results to fix the holomorphic ambiguity. For local 
$\projective^2$, we obtain in the above conventions
\begin{equation}
f^{(1,1)} = \frac {\ii}{24} (-z)^{1/2}
\end{equation}

There is one non-orientable diagram that contributes to the amplitude at 
$\chi=1$. Its holomorphic anomaly equation is solved by
\begin{equation}
\begin{split}
\del_\ib \K 11 &= \frac 12 \Cu_\ib^{jk} \Delta_{jk} - \klein_j \Delta_\ib^j  \\
&= \del_\ib\bigl(\frac 12 \Su^{jk} \Delta_{jk} - \klein_j \Delta^j\bigr)
+ \frac 12 \Su^{jk} C_{jkl}\Delta^l_\ib +\Bigl(\frac 12 C_{jkl} \Cu_\ib^{kl} - G_{\ib j}\Bigr)
\Delta^j \\
&= \del_\ib \bigl(\frac 12 \Su^{jk} \Delta_{jk} - \klein_j\Delta^j +
\frac 12 C_{jkl} \Su^{kl} \Delta^j - \Delta\bigr) 
\end{split}
\end{equation}
where $C^P_{ijk} = C_{ijl} P^l_k$, \cf., \eqref{asense}. Specializing to
our one-parameter models, we obtain
\begin{equation}
\eqlabel{omit}
\K 11 = -\klein_t\Delta^t - \Delta + f^{(1,1)_k}
\end{equation}
For local $\projective^2$, we find that we reproduce the localization results
of table \ref{GW2real} for $f^{(1,1)_k}=0$. For the quintic and the bicubic,
we find that the localization results are also reproduced, but in fact by the 
first term, $-\klein_t\Delta^t$ in \eqref{omit} {\it alone}. We interpret this 
to say that since there is no observable that would distinguish between $\F 11$ 
and $\K 11$, only the sum $\F 11+\K 11$ can have a truly invariant meaning. 
(In the A-model, this mixing is possibly related to our generous treatment of
real torus fixed points.) Consequently, we attempt to fix the holomorphic
ambiguity only for the combination
\begin{equation}
\F 11 + \K 11 = - \F 10_t\Delta^t - \klein_t\Delta^t -\frac{\chi}{24} \Delta
+ f^{(1,1)} \,,
\end{equation}
by requiring vanishing of the integral invariants in low degree. We find
for the quintic
\begin{equation}
f^{(1,1)} = \frac{85}8 \ii \varpi_0(z) \sqrt{z}
\end{equation}
and for the bicubic,
\begin{equation}
f^{(1,1)} = \frac{27}8 \ii \varpi_0(z) \sqrt{z}
\end{equation}
The expansion
\begin{equation}
\ii (\F 03 + \F 11 + \K 11) = 2 \sum_{\topa{d\;{\rm odd}}{k\;{\rm odd}}}
\bigl(n^{(2,{\rm real})}_d - \frac 1{24} n^{(0,{\rm real})}_d\bigr)
q^{kd/2}
\end{equation}
with $n^{(0,{\rm real})}_d$ taken from \eqref{shende} then delivers
the integers shown in table \ref{genustwo}.

\begin{table}[ht]
\begin{center}
\begin{tabular}{|c|r|r|r|}
\hline
 $d$ & quintic  &   bicubic & local $\projective^2$ \\\hline
1   &  $0 $                  & 0               &  0 \\
3   &  $0 $                  & $0$         &  $0$ \\
5   &  $-55640$              & $-693 $      & $-10$  \\
7   &  $-159440655 $         & $-568557$        & $229$ \\
9  & $-387012696805 $        & $-328426623$       & $-4833$ \\
11  & $-878665820903170 $    & $-175272593346$    & $96823$\\
\hline
\end{tabular}
\caption{Genus $\hat g=2$ real BPS invariants $n^{(2,{\rm real})}_d$}
\label{genustwo}
\end{center}
\end{table}

\subsection{Further checks}

Let us first digress a bit on the systematics of the BPS expansion
that we have been using. We recall that we are working in the normalization 
\eqref{found} for the total topological amplitude,
\begin{equation}
\G \chi = \frac 12 \Bigl[ \Fc {g_\chi} + \sum \F gh + \sum \K gh \Bigr]
\end{equation}
(where $g_\chi\equiv \frac\chi 2+1$). According to \cite{gova}, the purely
closed string contribution to this sum admits a large volume expansion of 
the form
\begin{equation}
\eqlabel{govaexp}
\sum_g\tilde\lambda^{2g-2} \Fc g (t)  = 
\sum_{g,d,k} n^{(g)}_d \, \frac{1}{k} 
\Bigl(2\sin\frac {\tilde\lambda k}{2}\Bigr)^{2g-2}\, q^{kd}
\end{equation}
in which all $n^{(g)}_d$ (Gopakumar-Vafa invariants) are integer.
Our computations in the previous subsections indicate that the rest of the 
amplitude should be expanded as
\begin{equation}
\eqlabel{myexp}
\sum_\chi \lambda^\chi \ii^{\chi} \Bigl(\G\chi - \frac 12 \Fc {g_\chi}\Bigr)=
\sum_{\topa{\chi\equiv d\bmod 2}{k\;{\rm odd}}} 
n^{(\hat g,{\rm real})}_d \, \frac{1}{k}
\Bigl(2\sinh\frac{\lambda k}{2}\Bigr)^\chi \, q^{kd/2}
\end{equation}
and that the $n^{(\hat g,{\rm real})}_d$ (with $\hat g = \chi+1$) should again
all be integer. These integers should give an invariant ``count'' of the number 
of real curves of genus $\hat g$ and degree $d$. (This interpretation is tied 
to our putting the D-brane on top of the orientifold plane.) Note that the 
expansions \eqref{govaexp} and \eqref{myexp} can in the future be related by 
identifying $\tilde\lambda=\ii\lambda$, absorbing the $\ii$ into the definition 
of the topological amplitudes, and redefining the integral invariants by a 
sign.

Let us now make a few steps in the direction of extending the computations to
higher order in perturbation theory. The integration of the holomorphic
anomaly equations is straightforward, although the resulting expressions fairly
quickly become too lengthy to write down explicitly. Most powerful is the
polynomial algorithm of \cite{yayau}, whose extended form is described in 
\cite{alim,konishi}. This reduces the problem to finding the right boundary
conditions on the amplitudes in order to fix the holomorphic ambiguity. A lot
of progress has recently been made on this problem, including in the compact
case \cite{hkq}. It seems likely that a more detailed analysis of the results 
in the present paper will allow a better understanding of the boundary
conditions also in the extended case. For the time being, we will extract 
what we can from our localization results in the A-model, and then present
a few checks of the enumerative aspects of the $n^{(\hat g,{\rm real})}_d$.

The amplitude at order $\chi=2$ receives contributions from four different
worldsheet topologies,
\begin{equation}
2\G 2 - \Fc 2 = \F 04 + \F 12 + \K 12 + \K 20
\end{equation}
For local $\projective^2$, we have enough A-model data to completely fix the 
holomorphic ambiguity of the four individual amplitudes, but we will only
present the results for the total amplitude, which can be directly computed
from the total anomaly equation \eqref{identical}. The holomorphic ambiguity 
is given by
\begin{equation}
\frac{7}{64} z + \frac{7}{288} \frac{1}{(1-27 z)} - \frac{9}{128} 
\frac{1}{(1-27 z)^2}
\end{equation}
and the BPS expansion \eqref{myexp} then gives the integers in table
\ref{genusthree}.

\begin{table}[ht]
\begin{center}
\begin{tabular}{|c|r|}
\hline
 $d$ & local $\projective^2$ \\\hline
2   &  0 \\
4   &  $1$ \\
6   &  $-63$ \\
8   & $2826$ \\
10  & $-91309$ \\
12  & $2548446$\\
\hline
\end{tabular}
\caption{Genus $\hat g=3$ real BPS invariants $n^{(3,{\rm real})}_d$}
\label{genusthree}
\end{center}
\end{table}

We do not have enough data to fix the holomorphic ambiguity of $\G 2$
for either the quintic or the bicubic. For $\chi=3$, the localization
data from table \ref{GW4real} is consistent with the present understanding,
and for local $\projective^2$ determines the following integer invariants:
\begin{equation}
n^{(4,{\rm real})}_1=0\,,\qquad
n^{(4,{\rm real})}_3=0\,,\qquad
n^{(4,{\rm real})}_5=-6\,,\qquad
\end{equation}

Let us now turn to the checks on the BPS invariants related to their 
interpretation as enumerating real curves. Consider the complex curves of
degree $d$ and genus $\hat g$, and let us pretend for simplicity that there 
is indeed a finite number $n^{(\hat g)}_d$ of them. Complex conjugation acts 
on this finite set, and we are claiming that the appropriately counted number 
of fixed points is given by $n^{(\hat g,{\rm real})}_d$.  Since all other 
orbits have order two, we immediately conclude that we must have
\begin{equation}
n_d^{(\hat g,{\rm real})} \equiv n_d^{(\hat g)} \bmod 2
\end{equation}
We also have the implication
\begin{equation}
n^{(\hat g)}_d = 0 \Rightarrow n^{(\hat g,{\rm real})}_d = 0
\end{equation}
One can easily check that these constraints are satisfied for all numbers
that we have listed in the tables above.

We can also verify a few of the numbers in the above tables directly. In general,
we expect that more checks can be done by taking a suitable real section of
the computational scheme for Gopakumar-Vafa invariants initiated in \cite{kkv}
and developed in several subsequent works. We also expect a connection with
appropriately defined ``real Donaldson-Thomas'' \cite{mnop} and ``real 
Pandharipande-Thomas'' \cite{path1,path2} invariants.

The simplest Gopakumar-Vafa invariants to check are those associated to
smooth curves. For local $\projective^2$, these are curves of genus 
$\hat g= \frac{(d-1)(d-2)}{2}$, which are parametrized by a copy of
$\complex\projective^{\frac{(d+1)(d+2)}{2}-1}$. Their contribution to 
Gopakumar-Vafa theory is up to a sign simply the Euler characteristic of 
the projective space,
\begin{equation}
n_d^{((d-1)(d-2)/2)} = (-1)^\frac{d^2+3d}{2} \frac{(d+1)(d+2)}{2}
\end{equation}
Clearly now, real curves of the same genus and degree are simply parametrized
by the corresponding real projective space $\RP^{\frac{d^2+3d}{2}}$, and
it is natural to assume that their contribution will (up to a sign) again 
be given by the Euler character of the parameter space. The interesting 
case is when $\frac{d^2+3d}{2}$ is even, when the Euler character is $1$.
Taking account of the constraint \eqref{onlyfor}, we see that when $d\equiv 0$
or $1\bmod 4$, the real invariant should be
\begin{equation}
n^{((d-1)(d-2)/2,{\rm real})}_d = \pm 1
\end{equation}
This leads to a check of $n^{(0,{\rm real})}_1=-1$ in table
\ref{zeroquintic} and $n^{(3,{\rm real})}_4=1$ in table \ref{genusthree}.

We can also interpret $n^{(0,{\rm real})}_3$ along those lines. The corresponding 
complex invariant is according to \cite{kkv} given by the Euler character of
the universal curve $\calc$ over the parameter space $\complex\projective^9$ 
of plane cubic curves. $\calc$ is a $\CP^8$ fibration over $\CP^2$, and
the Euler characteristic is $n^{(0)}_3=\eul(\calc)=\eul(\CP^2)\cdot\eul(\CP^8)=27$.
In the real version, this simply yields
\begin{equation}
n^{(0,{\rm real})}_3 = \eul(\calc^{\rm real}) = \eul(\RP^2)\cdot \eul(\RP^8) = 1
\end{equation}
in agreement with table \ref{zeroquintic} (up to possibly a sign which we shall 
not attempt to justify).

As a less elementary computation, we can verify the invariants $n^{(1,{\rm real})}_4
=-n^{(2,{\rm real})}_5=693$ on the bicubic. The corresponding complex invariants 
also coincide, $n^{(1)}_4=n^{(2)}_5=5520393$, see ref.\ \cite{hkq}. This 
coincidence is similar to $n^{(0)}_2=n^{(1)}_3=609250$ on the quintic, and can 
be understood as follows. Let $C_4$ be a smooth degree $4$ genus $1$ curve 
contained in the bicubic. Such a curve spans a unique $\complex\projective^3\subset
\CP^5$. This $\CP^3$ meets the bicubic in a degree $9$ curve $C_9$, which must be 
reducible, with one component being $C_4$. the other component is a degree $5$ 
curve $C_5$ of genus $2$. Conversely, we can start from $C_5$ to obtain $C_4$.%
\footnote{I thank Sheldon Katz for clarifying this.} Thus, $n^{(1)}_4=n^{(2)}_5$ 
on the bicubic. The coincidence of the two invariants should be preserved over 
the reals (again, up to a sign), which is exactly as predicted! But we can in 
fact do even better. 

The BPS invariant $n^{(1)}_4=3721431625$ on the quintic was verified in \cite{es}. 
There are two contributions. The first comes from smooth elliptic quartics in all
$\CP^3$'s inside of $\CP^5$, and can be computed by localization on the corresponding 
relative Hilbert scheme. The second contribution comes from plane binodal quartics,
and is obtained by a more classical computation. On the bicubic, we have
a contribution only from the smooth quartics, because the planes meet the
bicubic in too low dimension. (We have used this fact in the previous paragraph.)
Thus, $n^{(1)}_4=5520393$ on the bicubic is given by a simple localization 
calculation, and taking a real section of it readily confirms 
$n^{(1,{\rm real})}_4=693$. The invariants $n^{(0,{\rm real})}_3$ can be computed
in the same manner, as already noted for the quintic in \cite{open}.

\section{Conclusions}
\label{conclusions}

In this paper, we have shown that the extended holomorphic anomaly equation of
\cite{openbcov} can also be used to compute topological string amplitudes on
Calabi-Yau orientifolds. We have verified that the results match those obtained
in the A-model by a computational prescription that can be understood as a
real version of the localization formulas on the moduli space of maps.
The success of these computations indicates that there is a
well-defined underlying Gromov-Witten theory. We have found hints that
the correct moduli spaces to define this theory in fact contain domain 
curves of varying topology, but fixed Euler character. The essential 
idea is to absorb boundary components that are homologically trivial 
in the target Lagrangian and shrink to zero size on the domain curve by 
smoothing the developing real node into a crosscap. For this to make 
sense in general, the homology class of the Lagrangian wrapped by 
the D-brane must be equal to that of the fixed point set of the 
anti-holomorphic involution defining the orientifold (O-plane). We 
argued that this phenomenon in Gromov-Witten theory should be interpreted 
physically as a manifestation of a ``tadpole cancellation condition'' in 
the topological string. 

From the formal point of view, the most interesting result is that the 
holomorphic anomaly equation for the {\it total} amplitude $\G\chi$ of 
the orientifold model simply {\it coincides} with the extended holomorphic 
anomaly equation of \cite{openbcov}. In the A-model, we have obtained 
a satisfactory BPS interpretation of the total amplitudes, and have 
verified several of these predictions from the point of view of real 
enumerative geometry.

Our work leaves several questions unaddressed. Let us mention some of them.
\nxt From the practical point of view, the most immediate problem is to understand
the behavior of the orientifold amplitudes at the special points in moduli space
other than large volume. The behavior at the conifold should be determined by a 
simple orientifold version of the usual closed string story \cite{ghva}, and is 
hence probably also related to orientifolds of $\hat c=1$ string at self-dual radius. 
We expect new effects at the orbifold point due to the ``tensionless domainwalls'' 
that appear on the worldvolume of the D-brane, as explained in \cite{open,openbcov}.
\nxt In the physical string, there is an intimate relation between tadpole
cancellation and the cancellation of gauge and gravitational anomalies
\cite{polcai}. It seems likely that there should be a similar connection also
in the topological string, which it would be interesting to understand this
better. 
\nxt As a hint in this direction, recall that the oriented one-loop topological
amplitudes are expressible in the B-model in terms of holomorphic Ray-Singer
torsions for the $\delbar$-operator coupled to the appropriate vector bundle 
\cite{bcov2}. Consequently, the non-orientable one-loop amplitudes (M\"obius 
strip and Klein bottle) are certainly related to Ray-Singer torsions twisted by 
duality, although we have not been able to locate a convenient reference in which 
such objects are studied. As a consequence of this connection, the one-loop 
amplitudes generally exhibit a ``gravitational anomaly'', \ie, an explicit 
dependence on the background K\"ahler metric \cite{klva,pewi}. For the torus 
amplitude, this anomaly reduces to a volume-dependent factor when one computes 
for the Ricci-flat metric on the Calabi-Yau. More generally, the explicit 
metric dependence can be eliminated by studying appropriate virtual bundles with
vanishing topological Chern classes \cite{schumacher}, in a way analogous
to our tadpole cancellation using anti-branes.\footnote{The cancellation 
of the metric dependence of the annulus amplitude for brane/anti-brane 
configurations was pointed out to me by Cumrun Vafa in November 2006.} 
\nxt Also related to the question of anomaly cancellation is the role
played by torsion charges, which do figure into tadpole and anomaly cancellation
in the physical string. It might be somewhat difficult to come up with
a workable example of this in the topological string.
\nxt Perhaps the most intriguing speculation arises in connection with 
the so-called wavefunction interpretation of the topological string 
partition function \cite{wittenwave}. In \cite{newa}, it was 
conjectured that the solutions of the extended holomorphic anomaly
equation as one varies the D-brane background should furnish a basis 
of Witten's Hilbert space $\calh_W$ that arises upon quantization of
the symplectic vector space $H^3(Y,\reals)$. In the context of orientifolds, 
it seems very likely that there will only be a finite number of allowed 
brane configurations satisfying tadpole cancellation. This is suggestive
of a {\it distinguished finite-dimensional} Hilbert space
$\calh_{\rm phys.}$ of physical states inside of $\calh_W$,
which would be a quite remarkable lesson of the topological 
string on compact Calabi-Yau manifolds.
\nxt Finally, we can also envisage some applications in the context of string
phenomenology. It is well-known that the topological string one-loop amplitudes 
are related to threshold corrections for gauge and gravitational couplings of an 
associated superstring compactification on the same Calabi-Yau manifold 
\cite{agnt,bcov2}. The one-loop amplitudes also enter the D-instanton
induced superpotentials in type I/II compactifications 
\cite{cvetic,haack,uranga,shamit}, see \cite{blumenhagen} for a recent review. 
The holomorphic anomaly equation should now allow a principled computation of 
these couplings for a general Calabi-Yau.

\begin{acknowledgments}
I would like to thank 
Sergei Gukov,
Simeon Hellerman,
Sheldon Katz,
Hirosi Ooguri,
Rahul Pandharipande, 
Jake Solomon,
and Cumrun Vafa for valuable discussions and communications.
This work was supported in part by the Roger Dashen Membership 
at the Institute for Advanced Study and by the NSF under grant number 
PHY-0503584.
\end{acknowledgments}

%\newpage

%\appendix

%\section{The Numbers}

%\newpage


\begin{thebibliography}{99}
\addcontentsline{toc}{section}{References}
\renewcommand{\itemsep}{-.2cm}
\colorlinksblue
\small

\bibitem{bcov1}
  M.~Bershadsky, S.~Cecotti, H.~Ooguri and C.~Vafa,
  ``Holomorphic anomalies in topological field theories,''
  Nucl.\ Phys.\  B {\bf 405}, 279 (1993)
  [\hepth{9302103}].
  %%CITATION = NUPHA,B405,279;%%

\bibitem{bcov2}
 M.~Bershadsky, S.~Cecotti, H.~Ooguri and C.~Vafa,
  ``Kodaira-Spencer theory of gravity and exact results for quantum string
  amplitudes,''
  Commun.\ Math.\ Phys.\  {\bf 165}, 311 (1994)
  [\hepth{9309140}].
  %%CITATION = CMPHA,165,311;%%

\bibitem{grsch}
  M.~B.~Green and J.~H.~Schwarz,
  ``Anomaly Cancellation In Supersymmetric D=10 Gauge Theory And Superstring
  Theory,''
  Phys.\ Lett.\  B {\bf 149}, 117 (1984).
  %%CITATION = PHLTA,B149,117;%%

\bibitem{polcai}
  J.~Polchinski and Y.~Cai,
  ``Consistency of Open Superstring Theories,''
  Nucl.\ Phys.\  B {\bf 296}, 91 (1988).
  %%CITATION = NUPHA,B296,91;%%

\bibitem{fms}
 D.~Friedan, E.~J.~Martinec and S.~H.~Shenker,
  ``Conformal Invariance, Supersymmetry And String Theory,''
  Nucl.\ Phys.\  B {\bf 271}, 93 (1986).
  %%CITATION = NUPHA,B271,93;%%

\bibitem{polgimon}
  E.~G.~Gimon and J.~Polchinski,
  ``Consistency Conditions for Orientifolds and D-Manifolds,''
  Phys.\ Rev.\  D {\bf 54}, 1667 (1996)
  [\hepth{9601038}].
  %%CITATION = PHRVA,D54,1667;%%

\bibitem{tst}
 E.~Witten,
  ``Topological Sigma Models,''
  Commun.\ Math.\ Phys.\  {\bf 118}, 411 (1988).
  %%CITATION = CMPHA,118,411;%%

\bibitem{openbcov}
 J.~Walcher,
  ``Extended Holomorphic Anomaly and Loop Amplitudes in Open Topological
  String,''
  \arxiv{0705.4098}{hep-th}.
  %%CITATION = ARXIV:0705.4098;%%

\bibitem{marcos1}
  M.~Marino,
  ``Open string amplitudes and large order behavior in topological string
  theory,''
  \hepth{0612127}.
  %%CITATION = HEP-TH/0612127;%%

\bibitem{emo}
 B.~Eynard, M.~Marino and N.~Orantin,
  ``Holomorphic anomaly and matrix models,''
  JHEP {\bf 0706}, 058 (2007)
  [\hepth{0702110}.
  %%CITATION = JHEPA,0706,058;%%

\bibitem{agnt2}
  I.~Antoniadis, K.~S.~Narain and T.~R.~Taylor,
  ``Open string topological amplitudes and gaugino masses,''
  Nucl.\ Phys.\  B {\bf 729}, 235 (2005)
  [\hepth{0507244}].
  %%CITATION = NUPHA,B729,235;%%

\bibitem{coy}
  P.~L.~H.~Cook, H.~Ooguri and J.~Yang,
  ``Comments on the Holomorphic Anomaly in Open Topological String Theory,''
  Phys.\ Lett.\  B {\bf 653}, 335 (2007)
  [\arxiv{0706.0511}{hep-th}].
  %%CITATION = PHLTA,B653,335;%%

\bibitem{bonelli}
  G.~Bonelli and A.~Tanzini,
  ``The holomorphic anomaly for open string moduli,''
  JHEP {\bf 0710}, 060 (2007)
  [\arxiv{0708.2627}{hep-th}].
  %%CITATION = JHEPA,0710,060;%%

\bibitem{alim}
  M.~Alim and J.~D.~Lange,
  ``Polynomial Structure of the (Open) Topological String Partition Function,''
  JHEP {\bf 0710}, 045 (2007)
  [\arxiv{0708.2886}{hep-th}].
  %%CITATION = JHEPA,0710,045;%%

\bibitem{konishi}
  Y.~Konishi and S.~Minabe,
  ``On solutions to Walcher's extended holomorphic anomaly equation,''
  \arxiv{0708.2898}{math.AG}.
  %%CITATION = ARXIV:0708.2898;%%

\bibitem{wittenwave}
  E.~Witten,
  ``Quantum background independence in string theory,'' 
  \hepth{9306122}.
  %%CITATION = HEP-TH/9306122;%%

\bibitem{integrable}
M.~Aganagic, R.~Dijkgraaf, A.~Klemm, M.~Marino and C.~Vafa,
``Topological strings and integrable hierarchies,''
Commun.\ Math.\ Phys.\  {\bf 261}, 451 (2006)
[\hepth{0312085}].
%%CITATION = CMPHA,261,451;%%

\bibitem{anv}
M.~Aganagic, A.~Neitzke and C.~Vafa,
``BPS microstates and the open topological string wave function,''
\hepth{0504054}.
%%CITATION = HEP-TH/0504054;%%

\bibitem{newa}
  A.~Neitzke and J.~Walcher,
  ``Background Independence and the Open Topological String Wavefunction,''
  \arxiv{0709.2390}{hep-th}|.
  %%CITATION = ARXIV:0709.2390;%%

\bibitem{oooy}
H.~Ooguri, Y.~Oz and Z.~Yin,
``D-branes on Calabi-Yau spaces and their mirrors,''
Nucl.\ Phys.\  B {\bf 477}, 407 (1996)
[\hepth{9606112}].
%%CITATION = NUPHA,B477,407;%%

\bibitem{gova}
  R.~Gopakumar and C.~Vafa, 
  ``M-theory and topological strings. I, II,'' 
 \hepth{9809187}, \hepth{9812127}.
  %%CITATION = HEP-TH/9812127;%%
  %%CITATION = HEP-TH/9809187;%%

\bibitem{oova}
 H.~Ooguri and C.~Vafa,
  ``Knot invariants and topological strings,''
  Nucl.\ Phys.\  B {\bf 577}, 419 (2000)
  [\hepth{9912123}].
  %%CITATION = NUPHA,B577,419;%%

\bibitem{kontsevich}
  M.~Kontsevich,
  ``Enumeration Of Rational Curves Via Torus Actions,''
  \hepth{9405035}.
  %%CITATION = HEP-TH/9405035;%%

\bibitem{katzliu}
  S.~H.~Katz and C.~C.~Liu,
  ``Enumerative Geometry of Stable Maps with Lagrangian Boundary Conditions and
  Multiple Covers of the Disc,''
  Adv.\ Theor.\ Math.\ Phys.\  {\bf 5}, 1 (2002)
  [\mathag{0103074}].
  %%CITATION = 00203,5,1;%%

\bibitem{grza}
  T.~Graber and E.~Zaslow,
  ``Open string Gromov-Witten invariants: Calculations and a mirror
  'theorem',''
  \hepth{0109075}.
  %%CITATION = HEP-TH/0109075;%%

\bibitem{mayr}
  P.~Mayr,
  ``Summing up open string instantons and N = 1 string amplitudes,''
  \hepth{0203237}.
  %%CITATION = HEP-TH/0203237;%%

\bibitem{diaconescu}
  D.~E.~Diaconescu, B.~Florea and A.~Misra,
  ``Orientifolds, unoriented instantons and localization,''
  JHEP {\bf 0307}, 041 (2003)
  [\hepth{0305021}].
  %%CITATION = JHEPA,0307,041;%%

\bibitem{bfm1}
  V.~Bouchard, B.~Florea and M.~Marino,
  ``Counting higher genus curves with crosscaps in Calabi-Yau orientifolds,''
  JHEP {\bf 0412}, 035 (2004)
  [\hepth{0405083}].
  %%CITATION = JHEPA,0412,035;%%

\bibitem{open}
J.~Walcher,
``Opening Mirror Symmetry on the Quintic,''
Comm.\ Math.\ Phys.\ {\bf 276} 671-689 (2007)
[\hepth{0605162}]
%%CITATION = HEP-TH/0605162;%%

\bibitem{psw}
R.~Pandharipande, J.~Solomon and J.~Walcher,
``Disk enumeration on the Quintic 3-fold,''
\math{0610901} 

\bibitem{lmv}
  J.~M.~F.~Labastida, M.~Marino and C.~Vafa,
  ``Knots, links and branes at large N,''
  JHEP {\bf 0011}, 007 (2000)
  [\hepth{0010102}].
  %%CITATION = JHEPA,0011,007;%%

\bibitem{sinhav}
  S.~Sinha and C.~Vafa,
  ``SO and Sp Chern-Simons at large N,''
  \hepth{0012136}.
  %%CITATION = HEP-TH/0012136;%%

\bibitem{aahv}
 B.~Acharya, M.~Aganagic, K.~Hori and C.~Vafa,
  ``Orientifolds, mirror symmetry and superpotentials,''
  \hepth{0202208}.
  %%CITATION = HEP-TH/0202208;%%

\bibitem{bfm2}
 V.~Bouchard, B.~Florea and M.~Marino,
  ``Topological open string amplitudes on orientifolds,''
  JHEP {\bf 0502}, 002 (2005)
  [\hepth{0411227}].
  %%CITATION = JHEPA,0502,002;%%

\bibitem{eyon}
  B.~Eynard and N.~Orantin,
  ``Invariants of algebraic curves and topological expansion,''
  \mathph{0702045}.
  %%CITATION = MATH-PH/0702045;%%

\bibitem{remodel}
  V.~Bouchard, A.~Klemm, M.~Marino and S.~Pasquetti,
  ``Remodeling the B-model,''
  \arxiv{0709.1453}{hep-th}.
  %%CITATION = ARXIV:0709.1453;%%

\bibitem{dlp}
  J.~Dai, R.~G.~Leigh and J.~Polchinski,
  ``New Connections Between String Theories,''
  Mod.\ Phys.\ Lett.\  A {\bf 4}, 2073 (1989).
  %%CITATION = MPLAE,A4,2073;%%

\bibitem{horava1}
 P.~Horava,
  ``Strings on World Sheet Orbifolds,''
  Nucl.\ Phys.\  B {\bf 327}, 461 (1989).
  %%CITATION = NUPHA,B327,461;%%

\bibitem{sagnotti}
  M.~Bianchi and A.~Sagnotti,
  ``On the systematics of open string theories,''
  Phys.\ Lett.\  B {\bf 247}, 517 (1990).
  %%CITATION = PHLTA,B247,517;%%

\bibitem{horava2}
  P.~Horava,
  ``Equivariant topological sigma models,''
  Nucl.\ Phys.\  B {\bf 418}, 571 (1994)
  [\hepth{9309124}].
  %%CITATION = NUPHA,B418,571;%%

\bibitem{brho}
  I.~Brunner and K.~Hori,
  ``Orientifolds and mirror symmetry,''
  JHEP {\bf 0411}, 005 (2004)
  [\hepth{0303135}].
  %%CITATION = JHEPA,0411,005;%%

\bibitem{howa}
 K.~Hori and J.~Walcher,
  ``D-brane categories for orientifolds: The Landau-Ginzburg case,''
  \hepth{0606179}.
  %%CITATION = HEP-TH/0606179;%%

\bibitem{mowa}
  D.~R.~Morrison and J.~Walcher,
  ``D-branes and Normal Functions,''
  \arxiv{0709.4028}{hep-th}.
  %%CITATION = ARXIV:0709.4028;%%

\bibitem{diaconescu2}
 D.~E.~Diaconescu, A.~Garcia-Raboso, R.~L.~Karp and K.~Sinha,
  ``D-brane superpotentials in Calabi-Yau orientifolds.''
  \hepth{0606180}.
  %%CITATION = HEP-TH/0606180;%%

\bibitem{weichold}
G.~Weichhold,
``\"Uber symmetrische Riemannsche Fl\"achen und die Periodizit\"atsmodulen der
zugeh\"origen Abelschen Normalintegrale erster Gattung,'' 
Diss.\ Leipzig. Schl\"omilch Z. XXVIII. 321-352 (1883).

\bibitem{busep}
P.~Buser, M.~Sepp\"al\"a
``Real structures of Teichmu\"uller spaces, Dehn twists, and
moduli spaces of real curves,''
Math.\ Z.\ 232, 547--558 (1999).

\bibitem{cdgp}
P.~Candelas, X.~C.~De La Ossa, P.~S.~Green and L.~Parkes,
``A Pair Of Calabi-Yau Manifolds As An Exactly Soluble Superconformal
Theory,''
Nucl.\ Phys.\ B {\bf 359}, 21 (1991).
%%CITATION = NUPHA,B359,21;%%

\bibitem{zinger}
A.~Zinger,
``The Reduced Genus-One Gromov-Witten Invariants of Calabi-Yau Hypersurfaces,''
 \arxiv{0705.2397}{math.AG}

\bibitem{jake}
J.~P.~Solomon
``Intersection theory on the moduli space of holomorphic curves with Lagrangian 
boundary conditions,''
\mathsg{0606429}

\bibitem{melissa}
C.-C.~M.~Liu,
``Moduli of J-Holomorphic Curves with Lagrangian Boundary Conditions and 
Open Gromov-Witten Invariants for an $S^1$-Equivariant Pair,'' 
\mathsg{0210257}

\bibitem{agva}
M.~Aganagic and C.~Vafa,
``Mirror symmetry, D-branes and counting holomorphic discs,''
\hepth{0012041}.
%%CITATION = HEP-TH/0012041;%%

\bibitem{akv}
 M.~Aganagic, A.~Klemm and C.~Vafa,
  ``Disk instantons, mirror symmetry and the duality web,''
  Z.\ Naturforsch.\  A {\bf 57}, 1 (2002)
  [\hepth{0105045}].
  %%CITATION = ZNTFA,A57,1;%%

\bibitem{bhhw}
  I.~Brunner, K.~Hori, K.~Hosomichi and J.~Walcher,
  ``Orientifolds of Gepner models,''
  JHEP {\bf 0702}, 001 (2007)
  [\hepth{0401137}].
  %%CITATION = JHEPA,0702,001;%%

\bibitem{raul}
 K.~Hori, K.~Hosomichi, D.~C.~Page, R.~Rabadan and J.~Walcher,
  ``Non-perturbative orientifold transitions at the conifold,''
  JHEP {\bf 0510}, 026 (2005)
  [\hepth{0506234}].
  %%CITATION = JHEPA,0510,026;%%

\bibitem{faber}
 C.~Faber,
''Algorithms for computing intersection numbers on moduli spaces of curves, 
with an application to the class of the locus of Jacobians,''
\alggeom{9706006}

\bibitem{grpa}
T.~Graber, R.~Pandharipande, 
``Localization of virtual classes,''
Invent.\ Math.\ {\bf 135}  (1999), 487--518,
[\mathag{9601010}]

\bibitem{klza}
  A.~Klemm and E.~Zaslow,
  ``Local mirror symmetry at higher genus,'' 
  \hepth{9906046}
  %%CITATION = HEP-TH/9906046;%%

\bibitem{newissues}
E.~Witten,
  ``New Issues In Manifolds Of SU(3) Holonomy,''
  Nucl.\ Phys.\  B {\bf 268}, 79 (1986).
  %%CITATION = NUPHA,B268,79;%%

\bibitem{wittencs}
 E.~Witten,
  ``Chern-Simons Gauge Theory As A String Theory,''
  Prog.\ Math.\  {\bf 133}, 637 (1995)
  [\hepth{9207094}].
  %%CITATION = PMTMA,133,637;%%

\bibitem{dofi}
 M.~R.~Douglas and B.~Fiol,
  ``D-branes and discrete torsion. II,''
  JHEP {\bf 0509} (2005) 053
  [\hepth{9903031}].
  %%CITATION = JHEPA,0509,053;%%

\bibitem{yayau}
 S.~Yamaguchi and S.~T.~Yau,
  ``Topological string partition functions as polynomials,''
  JHEP {\bf 0407}, 047 (2004)
  [\hepth{0406078}].
  %%CITATION = JHEPA,0407,047;%%

\bibitem{dave}
D.~R. Morrison, 
``Mirror symmetry and rational curves on quintic threefolds:
  a guide for mathematicians,'' 
J.\ Amer.\ Math.\ Soc.\ {\bf 6} (1993) 223--247,
  [\alggeom{9202004}].

\bibitem{lmw}
W.~Lerche, P.~Mayr and N.~Warner,
``Holomorphic N = 1 special geometry of open-closed type II strings,''
\hepth{0207259};
%%CITATION = HEP-TH 0207259;%%
``N = 1 special geometry, mixed Hodge variations and toric geometry,''
\hepth{0208039}.
%%CITATION = HEP-TH 0208039;%%

\bibitem{strings}
K.~Hori and J.~Walcher,
``D-branes from matrix factorizations,''
Talk at Strings '04, June 28--July 2 2004, Paris.
Comptes Rendus Physique {\bf 5}, 1061 (2004)
[\hepth{0409204}].
%%CITATION = HEP-TH 0409204;%%

\bibitem{shende}
V.~Shende, unpublished

\bibitem{lite}
A.~Libgober, J.~ Teitelbaum, 
``Lines on Calabi-Yau complete intersections, mirror symmetry, and Picard-Fuchs equations,''
Internat.\ Math.\ Res.\ Notices 1993, no. 1, 29--39. 

\bibitem{hkq}
 M.~x.~Huang, A.~Klemm and S.~Quackenbush,
  ``Topological String Theory on Compact Calabi-Yau: Modularity and Boundary
  Conditions,''
  \hepth{0612125}.
  %%CITATION = HEP-TH/0612125;%%

\bibitem{kkv}
  S.~H.~Katz, A.~Klemm and C.~Vafa,
  ``M-theory, topological strings and spinning black holes,''
  Adv.\ Theor.\ Math.\ Phys.\  {\bf 3}, 1445 (1999)
  [\hepth{9910181}].
  %%CITATION = 00203,3,1445;%%

\bibitem{mnop}
D.~Maulik, N.~Nekrasov, A.~Okounkov, R.~Pandharipande,
``Gromov-Witten theory and Donaldson-Thomas theory. I,II,''
Compos.\ Math.\  {\bf 142}  (2006),  no. 5, 1263--1285, 1286--1304. 
[\mathag{0312059},\mathag{0406092}]

\bibitem{path1}
R.~Pandharipande, R.~P.~Thomas,
``Curve counting via stable pairs in the derived category,''
 \arxiv{0707.2348}{math.AG};

\bibitem{path2}
 R.~Pandharipande, R.~P.~Thomas,
 ``Stable pairs and BPS invariants,''
 \arxiv{0711.3899}{math.AG}
  
\bibitem{es}
  G.~Ellingsrud and S.~Str{\o}mme,
  ``Bott's formula and enumerative geometry,''
  J.~AMS {\bf 9}, 175--193 (1996)
  [\alggeom{9411005}].

\bibitem{ghva}
  D.~Ghoshal and C.~Vafa,
  ``C = 1 String As The Topological Theory Of The Conifold,''
  Nucl.\ Phys.\  B {\bf 453}, 121 (1995)
  [\hepth{9506122}].
  %%CITATION = NUPHA,B453,121;%%

\bibitem{klva}
A.~Klemm and C.~Vafa, unpublished.

\bibitem{pewi}
V.~Pestun and E.~Witten,
  ``The Hitchin functionals and the topological B-model at one loop,''
  Lett.\ Math.\ Phys.\  {\bf 74}, 21 (2005)
  [\hepth{0503083}].
  %%CITATION = LMPHD,74,21;%%

\bibitem{schumacher}
I.~Biswas and G.~Schumacher,
``Determinant bundle, Quillen metric, and Petersson-Weil form on moduli spaces,''  
Geom.\ Funct.\ Anal.\  {\bf 9},  no. 2, 226--255 (1999)

\bibitem{agnt}
 I.~Antoniadis, E.~Gava, K.~S.~Narain and T.~R.~Taylor,
  ``Topological amplitudes in string theory,''
  Nucl.\ Phys.\  B {\bf 413}, 162 (1994)
  [\hepth{9307158}].
  %%CITATION = NUPHA,B413,162;%%

\bibitem{cvetic}
  R.~Blumenhagen, M.~Cvetic and T.~Weigand,
  ``Spacetime instanton corrections in 4D string vacua - the seesaw mechanism
  for D-brane models,''
  Nucl.\ Phys.\  B {\bf 771}, 113 (2007)
  [\hepth{0609191}].
  %%CITATION = NUPHA,B771,113;%%

\bibitem{haack}
 M.~Haack, D.~Krefl, D.~Lust, A.~Van Proeyen and M.~Zagermann,
  ``Gaugino condensates and D-terms from D7-branes,''
  JHEP {\bf 0701}, 078 (2007)
  [\hepth{0609211}].
  %%CITATION = JHEPA,0701,078;%%

\bibitem{uranga}
  L.~E.~Ibanez and A.~M.~Uranga,
  %``Neutrino Majorana masses from string theory instanton effects,''
  JHEP {\bf 0703}, 052 (2007)
  [\hepth{0609213}].
  %%CITATION = JHEPA,0703,052;%%

\bibitem{shamit}
  B.~Florea, S.~Kachru, J.~McGreevy and N.~Saulina,
  ``Stringy instantons and quiver gauge theories,''
  JHEP {\bf 0705}, 024 (2007)
  [\hepth{0610003}].
  %%CITATION = JHEPA,0705,024;%%

\bibitem{blumenhagen}
  N.~Akerblom, R.~Blumenhagen, D.~L\"ust, M.~Schmidt-Sommerfeld,
``D-brane Instantons in 4D Supersymmetric String Vacua,''
  \arxiv{0712.1793}{hep-th}


%references

\end{thebibliography}
\end{document}